\documentclass{aa}

\usepackage{graphicx}
\usepackage{xcolor}%

\usepackage{txfonts}
\usepackage{hyperref}

\newcommand{\mdot}{\mbox{$\dot{M}$}}
\newcommand{\myr}{\mbox{${\rm M}_\odot\,{\rm yr}^{-1}$}}

\newcommand{\xmm}{{XMM-Newton}}

\def \qwr {HD\,45166}

\begin{document}



\title{X-ray and radio data obtained by XMM-Newton and VLA constrain the stellar wind of the magnetic quasi-Wolf-Rayet star in HD45166 }
\titlerunning{X-ray and radio observations of HD45166}
\authorrunning{Leto, Oskinova et al.}

\author{P.~Leto~\inst{1}, L.M.~Oskinova~\inst{2}, T.~Shenar~\inst{3}, G.A.~Wade~\inst{4}, S.~Owocki~\inst{5}, C.S.~Buemi~\inst{1}, R.~Ignace~\inst{6}, C.~Trigilio~\inst{1}, G.~Umana~\inst{1}, A.~ud-Doula~\inst{7}, H.~Todt~\inst{2}, W.-R.~Hamann~\inst{2}
          }

\institute{
Osservatorio Astrofisico di Catania, INAF, Via S. Sofia 78, I-95123 Catania, Italy\\
\email{paolo.leto@inaf.it}
\and              
Institute for Physics and Astronomy, University Potsdam, Karl-Liebknecht-Str. 24/25, D-14476 Potsdam, Germany\\
\email{lida@astro.physik.uni-potsdam.de}
\and              
The School of Physics and Astronomy, Tel Aviv University, 6997801 Tel Aviv, Israel
\and              
Department of Physics and Space Science, Royal Military College of Canada, PO Box 17000, Kingston, ON K7K 7B4, Canada
\and
Department of Physics and Astronomy, University of Delaware, 217 Sharp Lab, Newark, DE 19716, USA 
\and
Department of Physics \& Astronomy, East Tennessee State University, 173 Sherrod Drive, Johnson City, TN 37614, USA
\and
Penn State Scranton, Pennsylvania State University, 120 Ridge View Drive, Dunmore, PA 18512, USA
}




 \abstract
{
Recently, a powerful magnetic field was discovered in the hot helium star classified as a quasi-Wolf-Rayet (qWR) star of $\sim 2$\,M$_\odot$, member of the \qwr\ system. Upon its explosion as a core-collapse supernova, it is expected to produce a strongly magnetic neutron star -- a magnetar. Among the key parameters governing the pre-supernova evolution is the amount of mass lost via stellar wind. However, the magnetic nature of this helium star is expected to affect its stellar wind making the estimation of the wind parameters uncertain.
}
{
We report the first observations of \qwr\ in X-rays with the XMM-Newton telescope and in radio with the VLA interferometer array. By placing the observation results in a theoretical framework, we aim to provide a reliable estimate of the wind strength of the magnetic qWR star.
}
{
The X-ray properties are explained in the framework of the magnetically-confined wind shock (MCWS) scenario, and the semi-analytic XADM model is applied to reproduce the X-ray emission. The thermal radio emission of the wind and its absorption effect on possible gyro-synchrotron emission from the underlying dipolar magnetosphere, sampled in 3D, are computed by integrating the radiative transfer equation.
}
{
We did not detect radio emissions, this enabled us to set sensitive upper limits on the radio luminosity. The magnetic qWR star is a slow rotator, comparison with models reveals that the possible acceleration mechanisms occurring within its dynamical magnetosphere are not as efficient as in fast-rotating magnetic ApBp-type stars. In contrast, the system is detected in X-rays with $\log(L_{\rm X}/L_{\rm bol})\sim{-5.6}$. Using suitable models, we constrain the mass lost from this magnetic quasi-Wolf-Rayet star as $\dot{M}\approx 3\times10^{-10}$\,\myr.
}
{
This novel empirical estimate of the mass-loss rate in a $\sim2$\,M$_\odot$ helium star confirms that it maintains super-Chandrasekhar mass till collapse and can produce a magnetar as its final evolutionary product.
}

\keywords{X-rays: stars -- Radio continuum: stars -- stars: Wolf-Rayet -- stars: winds, outflows -- stars: mass-loss -- stars: magnetars}

   \maketitle
%

\section{Introduction}
\label{sec:introduction}

The binary system \object{HD 45166} consists of a hot helium star and a B7\,V-type star \citep{Willis1983} bound with an orbital period of 22\,yr \citep{Shenar2023}. The optical spectrum of the helium star is dominated by emission lines of carbon, nitrogen, and oxygen in high ionization states, hence warranting its Wolf-Rayet (WR) spectral classification \citep{Anger1933, Neubauer1948, Hiltner1966}. The relatively low luminosity of the helium star compared to classical WR stars, in combination with its uniquely narrow emission lines and unusual abundance pattern, gave rise to its designation as a quasi-Wolf-Rayet (qWR) star \citep{vanBlerkom1978}. The mass of the qWR star was originally estimated at $0.5\,M_\odot$ \citep{Willis1983}, then raised to $\approx 4\,M_\odot$ \citep{Groh2008}, and finally revised as $M_\ast = 2.03\pm0.44$\,M$_\odot$ \citep{Shenar2023}. While this mass is at the lower limit needed for core-collapse \citep{Woosley1995} the final fate of the qWR component is uncertain, and whether or not the star will produce a supernova (SN) will sensitively depend on the amount of mass it will lose until core-collapse.

\begin{table*}[!htb]
\centering
\caption{Parameters of the magnetic qWR star in the HD~45166 system.} 
\begin{tabular}{lccc}
\hline  \hline
Parameter &Unit &  Value & Reference  \\
\hline
\vspace{-3mm}\\ 
Distance: $D$ & parsec & $991^{+38}_{-33}$ & \cite{Bailer-Jones2021}  \\
\vspace{-3mm}\\ 

Bolometric luminosity: $\log (L/L_\odot)$ & - &  $3.83\pm0.05$  &\cite{Shenar2023}\\

Reddening: $E_{B - V}\,$ & magnitude & $0.210\pm0.010$  & \cite{Shenar2023} \\

Temperature: $T_*$ & kK     & $56.0\pm6.0$ &\cite{Shenar2023}  \\

Mass: $M_*$ & M$_\odot$ &  $2.03\pm 0.44$ &\cite{Shenar2023}  \\ 

Radius: $R_*$ & $R_\odot$  &  $0.88\pm 0.16$ & \cite{Shenar2023} \\

Rotation period: $P_{\rm rot}$ &  day &  $124.8\pm 0.2$ &  \cite{Shenar2023} \\

Average magnetic field: $\langle B \rangle$ & kG  & $ 43.0\pm2.5$  & \cite{Shenar2023}\\

\hline

\end{tabular}

\label{tab:qwr}
\end{table*}

The characteristics of the helium star in \qwr\  make it a prototype for super-Chandrasekhar mass stars stripped of their outer hydrogen layers. Helium stars of such masses are generally thought to form via interactions among components in close binaries \citep{Podsi1992, Yung2024}. This evolutionary channel can lead to the formation of helium-rich stars significantly less massive than the classical WR stars (i.e.\,$\lesssim 10$\,M$_\odot$) \citep{Hamann2019, Sander2019, Shenar2020}. Binary evolution models predict that stripped stars are hot and numerous; as such, they could be among the dominant ionizing sources in star-forming galaxies and may represent the bulk of progenitors of stripped supernovae \citep{Dionne2006, Goetberg2018, Doughty2021}. Finding super-Chandrasekhar-mass helium stars stripped by binary interactions turned out to be difficult \citep{Ramachandran2023, Drout2023,Gilkis2023}, and over a long time, the qWR star in the \qwr\ system was the only well-studied candidate. 

Of special interest is the mass-loss rate of these objects, which determines the properties of the pre-collapse core. Radiation fields of hot, OB- and WR-type, stars drive stellar winds \citep{Castor1975}. Previous analyses of the optical spectrum of the qWR star by means of latitude-dependent wind models derived a mass-loss rate of $\dot{M} \approx 2 \times 10^{-7}\,M_\odot\,{\rm yr}^{-1}$ (with: equatorial wind velocity $V_{\infty}^{\rm eq}\approx 425$\,km\,s$^{-1}$; polar terminal wind velocity $V_{\infty}^{\rm pole}\approx 1200$\,km\,s$^{-1}$; \citealp{Willis1983, Willis1989, Groh2008}). These values are significantly different from the theoretical predictions. \citet{Vink2017} explored the parameter space characterizing winds from helium stars. The mass of the helium star (or equivalently its luminosity given the mass-luminosity relation) has been varied in the range from 60\,M$_{\odot}$ to 0.6 solar masses, with a fine tuning in the range 2--20\,M$_{\odot}$. Once assigned metallicity and mass, the Monte Carlo modeling approach performed by \citet{Vink2017} provides both the mass-loss rate and the terminal wind velocity. In particular, the mass-loss rate predicted for a 2\,M$_\odot$ star at Solar metallicity is $\dot{M}_{\rm th} \approx 4 \times 10^{-9}$ M$_{\odot}$\,yr$^{-1}$, while the predicted wind velocity is $V_{\infty}^{\rm th} \approx 2650$\,km\,s$^{-1}$. 

The striking difference between the empirically measured mass-loss rate and wind velocity of the qWR star and the theoretical predictions have been noticed already by \citet{Vink2017}. However, the recent discovery of strong magnetic field on the qWR star suggests that its wind is strongly affected by magnetic field and calls for new approaches in evaluating its stellar wind parameters. In this paper we address this problem using new X-ray and radio observations.

The qWR possesses an exceptionally strong magnetic field of $\langle B \rangle \sim 43$\,kG. This breaks the record held by Babcock's Star  ($\sim 34$\,kG; \citealp{Babcock1960}) making the qWR the most strongly magnetic non-degenerate star known. The qWR is likely a merger product in an initially triple system, with the non-magnetic B7\,V component being an original tertiary, and is predicted to end its life as a strongly magnetic neutron star: a magnetar \citep{Shenar2023}.

The high magnetic field of the qWR star plays a major role in regulating its mass-loss. The magnetic field affects the free radial flow of the stellar wind up to a certain distance, named Alfv\'en radius ($R_{\rm A}$), leading to a so-called confined wind structure. Hence, the spectrum of the qWR star forms in its magnetosphere, which must be taken into account in empirical mass-loss rate measurements.

\begin{table*}
\begin{center}
\caption[ ]{Log of the VLA Observations of \qwr. Array Config C. Code: 22B-314.}
\label{tab:vla_obs}
\footnotesize
\begin{tabular}{l c c c c c c c}
\hline
\hline
Date-OBS   &UTC             &T. on Sou.  & $\nu$  &$\Delta\nu$ & RMS            &{FWHM}  & {PA}\\
                   &                    &(minutes)   & (GHz)  & (GHz)& ($\mu$Jy/beam) &{($^{\prime\prime} \times ^{\prime\prime}$)} &{(deg)}\\
\hline
2022 Nov 9 &13:13:47.5   &$19$               &5.5      &2                 &  $\approx 8$  &{$4.88 \times 3.69$}  & {50.57} \\
2022 Nov 9 &13:38:40.5   &$19$               &9         &2                 &  $\approx 7$  &{$3.60 \times 2.18$}  & {54.22} \\
2022 Nov 16 &13:24:40.5   &$26$               &15       &6                 &  $\approx 5$  &{$2.26 \times 1.37$}  & {54.88} \\
\hline
\end{tabular}
\end{center}
\end{table*}

The magnetic star in the \qwr\ system is a very slow rotator, its rotation period exceeds 100 days (Table\,\ref{tab:qwr}). This is a suitable condition for originating a dynamical magnetosphere (DM) which commonly characterizes the slowly rotating massive magnetic stars with a radiatively driven stellar wind \citep{Petit2013}. Furthermore, the combined presence of wind and magnetic field allows to place the qWR in the context of other hot massive magnetic stars such as ApBp-type stars which are, in general, X-ray and radio sources \citep{Drake1987, Drake1994, Linsky1992, Leone_etal1994,Osk2011,Naze2014}. Production of X-rays in magnetic hot stars is usually explained by the cooling of a fraction of their stellar winds, heated to a few million degrees by strong shocks resulting from the collision of streams confined by a magnetic field (magnetically confined wind shock, MCWS, model) \citep{Babel1997,udD2016}. This helps to explain why some magnetic stars are more X-ray luminous compared to their non-magnetic counterparts \citep{Osk2011, Naze2014}.

Magnetohydrodynamic (MHD) simulations predict that the continuous supply of wind material to the magnetospheres in slowly-rotating stars is balanced by the plasma infall back onto the stellar surface \citep{ud-doula2002,ud-doula2008, ud-doula2013}. Beyond the Alfv\'en radius the breaking of the magnetic field lines should lead to particle acceleration \citep{Usov_Melrose1992}. Fast non-thermal electrons power both incoherent non-thermal gyro-synchrotron and coherent auroral radio emission \citep{Trigilio2000,Das_etal2022} and, in some cases, also non-thermal X-rays of auroral origin \citep{Leto2017, Robrade2018}.

Thus, radio and X-ray observations are excellent probes of conditions in stellar magnetospheres.
To gain insight into the wind properties of the qWR star and its magnetosphere we obtained new observations in X-ray and radio domains.
In Sect.\,\ref{sec:observations} we describe the observations and data reduction.
In Sect.\,\ref{sec:obs_results} we report the observational results directly derived from the analysis of the X-ray (Sect.\,\ref{sec:Xray_diagn}) and radio (Sect.\,\ref{sec:radio_diagn}).
In Sect.\,\ref{sec:xraytermo} we provide the theoretically expected X-ray spectrum;
in Sect.\,\ref{sec:wind_and_magfield} we describe how the stellar wind is affected by its strong magnetic field. 
The physical mechanism affecting the radio emission are explored in Sect.\,\ref{sec:nonterm_radio};
in Sect.\,\ref{sec:accel_mechanism} we discuss the possible non-thermal electron production and, after taking into account the frequency dependent absorption effect due to the large-scale distributed wind plasma, the upper limit of the corresponding non-thermal radio emission has been calculated in Sect.\,\ref{sec:nontermspec}, 
finally, in Sect.\,\ref{sec:discussion}, we place our results in a general framework that includes the O-type magnetic stars.
In Sect.\,\ref{sec:conclusions} we present our conclusions and discuss how the forthcoming sensitive  radio facilities may be useful for the science case discussed in this paper.

\section{Observations and data reduction}
\label{sec:observations}

\subsection{Radio}
\label{sec:radio_data}

The radio observations were performed at the Very Large Array (VLA) National Radio Astronomy Observatory in November 2022, the observing log is reported in Table~\ref{tab:vla_obs}. The observations were conducted in three bands: the C-band, centered at $\nu=5.5$ GHz; the X-band, centered at $\nu=9$ GHz; and the Ku-band, centered at $\nu=15$ GHz. For the C and X bands the adopted hardware setup allowed to observe a bandwidth of 2 GHz width (8-bit digital samplers), whereas the adopted setup for the Ku-band observations allowed recovering flux within a wider spectral range, bandpass of 6 GHz using the 3-bit digital samplers.

To calibrate the flux scale and the receiver's response within the spectral range covered by the VLA receivers (bandpass calibration), the radio galaxy 3C286 (1331+305) was observed as primary calibrator in each observing band. To calibrate the amplitude and phase of the complex gain, the standard VLA calibrator J0613+1306 was cyclically observed for each band during the observing scans. J0613+1306 is a point-like radio source located close to the sky position of \qwr\ (about $6^{\circ}$ away) with an almost flat radio emission level ($\approx 0.5$ Jy) in the range 5--15 GHz. The observations have been processed through the VLA Calibration Pipeline  (version 2022.2.0.64), which is designed to handle Stokes I continuum data, operating within the Common Astronomy Software Applications  ({\sc casa}) package (release  6.4.1). Images of the sky region centered at the target position have been obtained using the task {\sc tclean} (number of Taylor terms 2, number of clean iterations 40000).

\qwr\ is located in a sky region not contaminated by strong radio sources and the radio maps of the total intensity (Stokes I) show no issues. In fact, we measured low noise levels in each band ($\mathrm{RMS} < 8$\,$\mu$Jy/beam, see Table~\ref{tab:vla_obs}), which nearly coincide with the nominal levels expected in each band, corresponding to the bandpass adopted and the available times on source, as given by the VLA exposure calculator\footnote{\url{https://obs.vla.nrao.edu/ect/}}, confirming the goodness of the VLA observations. Despite the high-quality radio measurements, \qwr\ is not detected. 

\begin{figure}[h]
\centering
\includegraphics[width=0.98\columnwidth]{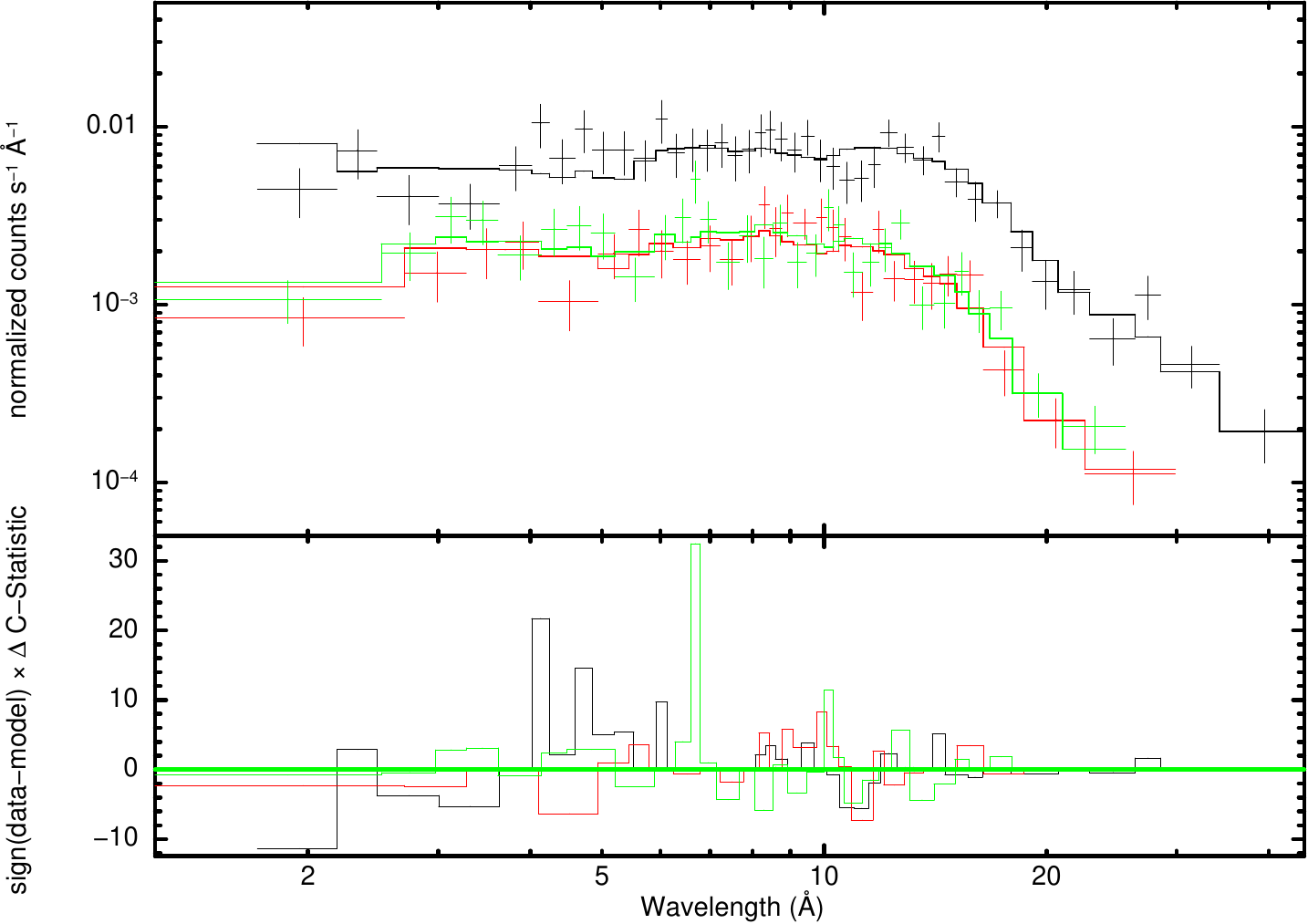}
\vspace{2mm}
\includegraphics[width=.98\columnwidth]{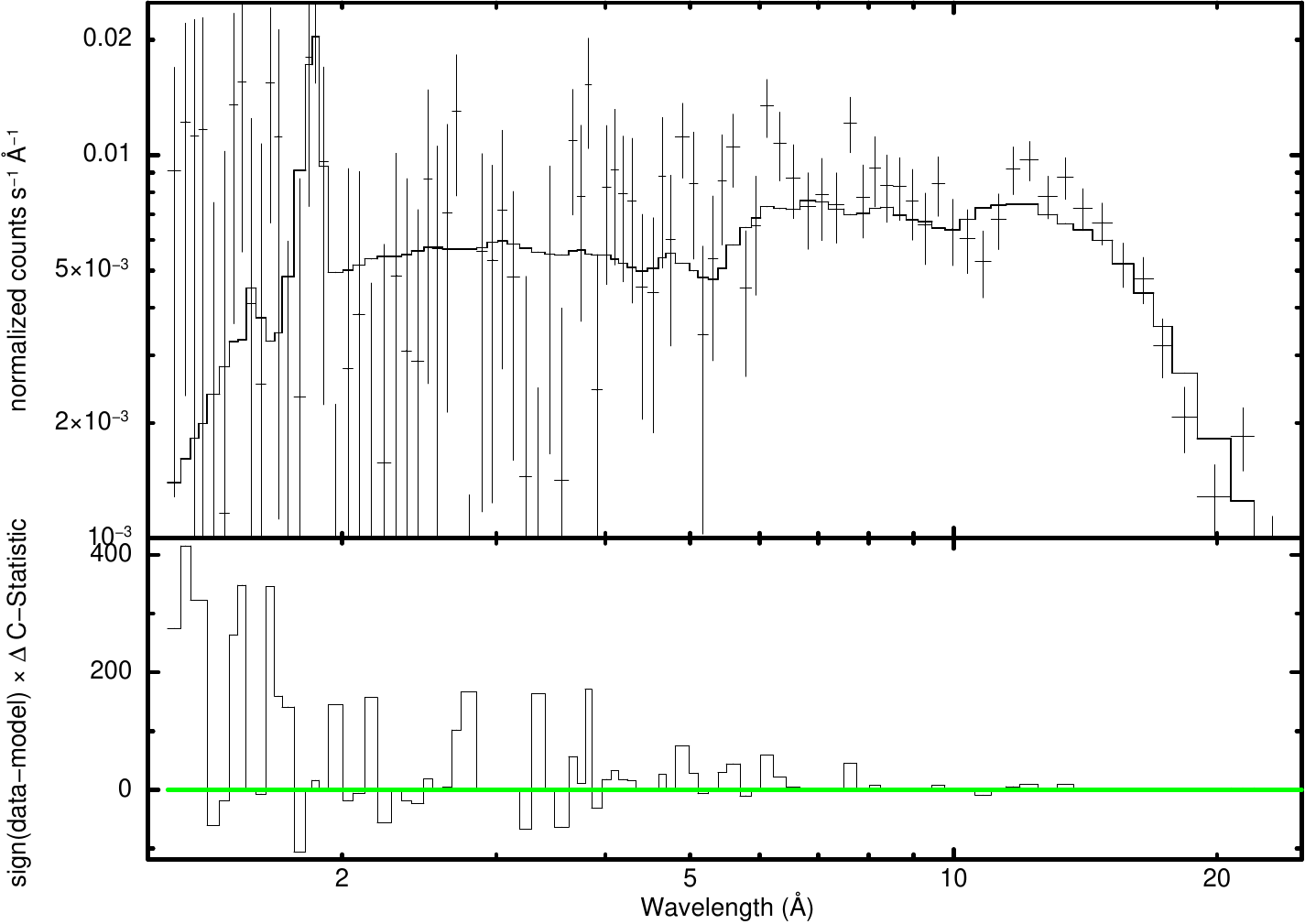}
\caption{Top panel: \xmm\ pn (upper black crosses), and MOS1 and MOS2 (lower green and red crosses) spectra of \qwr\ where a suitable number of instrumental channels are binned, but not more than 300, to achieve at least a 3$\sigma$ detection. Bottom panel: \xmm\ unbinned EPIC pn spectra corrected for the instrument response (black crosses) of \qwr, error bars corresponding to 1$\sigma$. The best-fit model is a combination of three thermal {\it vapec} models. 
}
\label{fig:spec1}
\end{figure}

\subsection{X-rays}
\label{sec:xray_data}

X-ray data were acquired with the X-Ray Multi-Mirror Mission (\xmm) of the European Space Agency (ESA). \xmm\ has three X-ray telescopes that illuminate five different instruments, which always operate simultaneously and independently. The useful data were obtained with the three focal instruments: MOS1, MOS2, and pn, which together form the European Photon Imaging Camera (EPIC). The EPIC instruments have a broad wavelength coverage of $1.2-60$\,\AA\ and allow low-resolution spectroscopy with ($E/\Delta E \approx 20-50$). Throughout this paper, X-ray fluxes and luminosities are given for the full energy band. 

The \xmm\ observations of \qwr\ were carried {out} on 2022-09-11 with a total duration $\sim 20$\,ks (ObsID 0900520101). All three EPIC cameras were operated in the standard, full-frame mode. The ``medium'' UV filter was used for MOS cameras, while the ``thin''  filter was used for the pn. The observations were affected by episodes of high background. After rejecting these time intervals, the cumulative useful exposure time was $\approx 6$\,ks for the EPIC pn and $\approx 12$\,ks for the EPIC MOS cameras. No significant source variability is seen during these exposure times. The data were analyzed using the  \xmm\ data analysis package SAS\footnote{\url{www.cosmos.esa.int/web/xmm-newton/what-is-sas}}. The customary and the pipeline-reduced data are consistent.

Contrary to the radio, \qwr\ is clearly detected in all \xmm\ cameras. The total EPIC count rate from the isolated X-ray source at the position of \qwr\ is $0.23 \pm 0.01$\,s$^{-1}$. The X-ray spectra and light curves of \qwr\ were extracted using standard procedures from a region with a radius $\approx 20^{\prime\prime}$. The background area was chosen to be nearby the star and free of X-ray sources. There are $\approx 1100$ spectral data counts registered by the pn camera, while MOS cameras registered $\approx 1040$ spectral counts.

To analyze the X-ray spectra of \qwr\ we used the standard spectral fitting software {\sc xspec} \citep{xspec}. The abundances were set to the \qwr\ abundances \citep{Shenar2023} using the method outlined in \citet{Osk2012, Osk2020}. In all spectral models, the absorption in the interstellar medium is included using the {\it tbabs} model \citep{tbabs}. Distance and reddening of \qwr\ are reported in Table\,\ref{tab:qwr}.

\begin{figure}[h!]
\centering
\includegraphics[width=0.98\columnwidth]{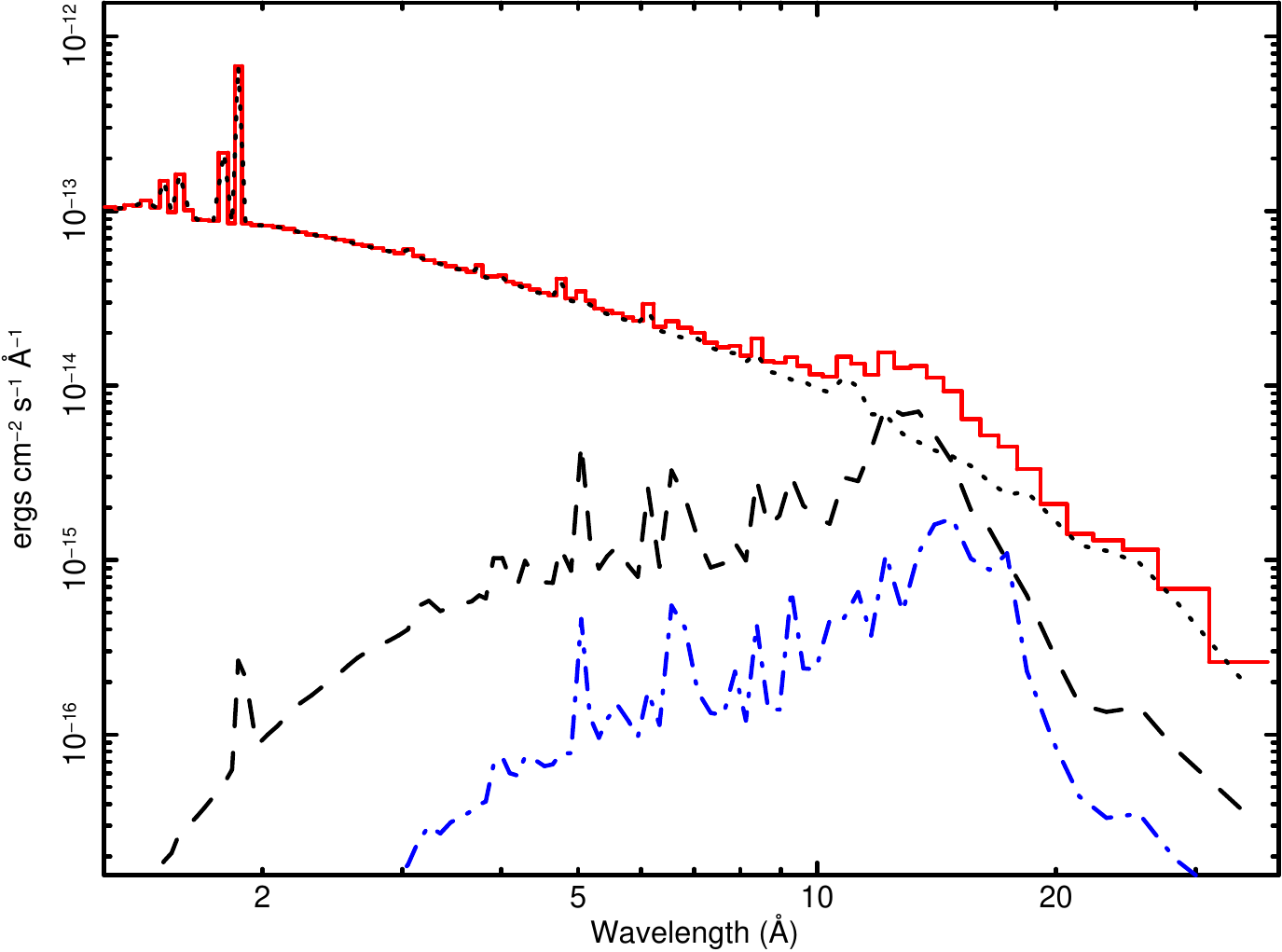}
\caption{
Contributions from individual model components to the best-fit spectral models. {\it Left panel:} three temperature thermal model. The three lines (blue dashed-dotted, black dashed, black dotted) show spectral models corresponding to individual components (see Table\,\ref{tab:spec}). The red solid line shows the best fit combined model. 
}
\label{fig:spec-uf}
\end{figure}

\section{Results}
\label{sec:obs_results}

\subsection{X-ray diagnostic of the hot plasma}
\label{sec:Xray_diagn}

The \qwr\ system contains a non-magnetic B7\,V-type star. Late B-type non-magnetic stars do not emit X-rays \citep{Stelzer2005,Evans2011_lateB_no_Xray_sources}, therefore we attribute all X-rays detected in \qwr\ only to its helium star companion. 

X-ray spectra of \qwr\ are shown in top panel of Fig.\,\ref{fig:spec1}. The corresponding unbinned EPIC pn spectrum of \qwr\ is shown in bottom panel of Fig.\,\ref{fig:spec1}, where a prominent emission line at $\lambda \approx 1.87$\,\AA\ is clearly seen. This line belongs to the He-like Fe\,{\sc xxv} ion which has a peak emissivity at $kT=5.4$\,keV ($T\approx 60$\,MK).

\begin{table}
\begin{center}
\caption[ ]{
X-ray spectral parameters derived from the \xmm\ EPIC observations of \qwr\ assuming a three-temperature thermal plasma ({\it vapec}) model and a thermal ({\it vapec}) plus a power-law model, both models corrected for the interstellar absorption using the {\it tbabs} model.  The errors are at 68\%\ confidence level. The abundances are set to those deduced from the spectral analysis of optical and UV spectra of \qwr\ \cite{Shenar2023}. The corresponding spectral fits are shown in Fig.~\ref{fig:spec1}. The contributions of each spectral component are shown in Fig.~\ref{fig:spec-uf}.
}
\label{tab:spec}
\footnotesize
\begin{tabular}[]{lc}
\hline
\hline

\multicolumn{2}{l}{Three temperature thermal model}\\
\hline

\vspace{-2mm}\\ 

$N_{\mathrm H}$ [$10^{21}$ cm$^{-2}$] & $1.2 \pm 0.3$ \\
$kT_1$ [keV]                          & $0.5  \pm 0.3$  \\
$EM_1$ [$10^{53}$ cm$^{-3}$]          & $1.1 \pm 0.6 $     \\
$kT_2$ [keV]                          & $0.96  \pm 0.21$    \\
$EM_2$ [$10^{53}$ cm$^{-3}$]          & $1.3 \pm 0.6 $   \\
$kT_3$ [keV]                          & $5.76 \pm 0.95$    \\
$EM_3$ [$10^{53}$ cm$^{-3}$]          & $20.2 \pm 1.21$   \\
Flux [$10^{-13}$ erg cm$^{-2}$ s$^{-1}$]  & $4.1 \pm 0.2$  \\
$L_{\mathrm X}^{\mathrm a}$ [erg s$^{-1}$]  & $(6.1\pm 0.2) \times 10^{31}$   \\

{$\log (L_{\rm X}/ L_{\mathrm {bol}})$} &    {$-5.63\pm0.05$}  \\

Fit statistic: $\chi^2$ with 96 d.o.f. &  116 \\

\vspace{-2mm}\\ 

\hline

\end{tabular}
\begin{list}{}{}
\item[$^{\mathrm{a}}$] intrinsic (dereddened) X-ray luminosity in the 0.2--12 keV energy band at the distance of $991$\,pc
\end{list}
\end{center}
\end{table}

To determine the physical conditions of the plasma, we fitted the observed spectra using the Astrophysical Plasma Emission Code {\it ``apec''} \citep{apec}, that is a state-of-the-art code able to model a thermal optically thin plasma in collisional equilibrium. We used the {\it vapec} version of the code which allows for non-solar abundances. The simulated spectrum superimposed to the observed one (binned and unbinned) is pictured in Fig.\,\ref{fig:spec1}.The corresponding model parameters are shown in Table\,\ref{tab:spec}. Applying the relation $N_{\mathrm {H, ISM}}= E_{B - V} \cdot 6.12 \times 10^{21}$\,cm$^{-2}$ \citep{Gudennavar2012}, the interstellar column density is $(1.28\pm0.06)\times 10^{21}$\,cm$^{-2}$. This is consistent with the $N_{\mathrm H}$ values inferred from the adopted X-ray model (see Table\,\ref{tab:spec}). Adopting the distance and the reddening from Table\,\ref{tab:qwr}, the X-ray luminosity of \qwr\ (Table\,\ref{tab:spec}) is $\log({L_{\rm X}\,[{\rm erg\,s}^{-1}]}) \approx 31.8$ corresponding to a ratio between X-ray and bolometric luminosity of $\log (L_{\mathrm X} / L_{\mathrm {bol}}) \approx -5.6$, i.e.\ among the highest ratios of known OB and WR-type stars without a compact binary companion \citep{Naze2014, Nebot2018}.

The combination of three temperature components is sufficient to fit the low-resolution X-ray spectrum of \qwr. The individual model components are shown in Fig.\,\ref{fig:spec-uf}. The lower temperature plasma components have $T\approx 6$ and 13\,MK, whereas the highest temperature plasma has $T\approx 70$\,MK ($kT\approx 5.7$\,keV), the largest emission measure ($EM$) among the three plasma components (Table\,\ref{tab:spec}), and is also responsible of the short wavelengths ($\lambda < 2$\,\AA) emission lines. To explain this high temperature by strong shocks invoking the Rankine-Hugoniot condition, where the temperature of the plasma after the shock is related to the pre-shock velocity as follows: $T  \approx 14\,\text{MK}  \times V_3^2$, where $V_3= V/(10^3 \,{\text {km\,s}}^{-1})$, a velocity jump of about $2200$\,km\,s$^{-1}$ is required. Therefore, if X-rays originate from shocks occurring in the magnetically confined wind, the three thermal components could be the tracers of the temperature range of the regions where the shocks dissipate energy.

\subsection{Radio diagnostic of the stellar wind}
\label{sec:radio_diagn}

Any ionized stellar wind also emits thermal radio. For a spherical non-magnetic wind, the radio spectrum is described by $S_{\nu} \propto \dot{M}^{4/3} \nu^{0.6}$  \citep{Wright1975,Panagia1975}, with the radio flux then providing a measure of the mass-loss rate. The full expression for calculating the theoretical spectrum of thermal radio emission from a stellar wind is given by \citet{Scuderi1998}:
\begin{align}
S_{\nu}=7.26 
\left(\frac{\nu}{{{10\,\mathrm{GHz}}}}\right)^{0.6}  
\left(\frac{T_{\mathrm{e}}}{10^4 \,{\mathrm K}}\right)^{0.1} 
\left(\frac{\dot M}{10^{-6} \,{\mathrm {M_{\odot} \, yr^{-1}}}}\right)^{4/3} \nonumber \\
\left(\frac{1.3 \, V_{\infty}}{100 \,{\mathrm {km \, s^{-1}}}}\right)^{-4/3} 
\left(\frac{d}{{10^3\,\mathrm{pc}}}\right)^{-2}  
\mathrm{~~mJy,} 
\label{eq:windspec1}
\end{align}

\noindent 
where the wind temperature ($T_{\mathrm{e}}$) is assumed to be 85\% of the effective stellar temperature. 

No radio emission is detected at the position of \qwr\ and only upper limits on flux in three radio bands could be determined. Once fixed the terminal wind velocity in the Eq.\,(\ref{eq:windspec1}), we can use the stringent upper limits on the fluxes in three radio bands (lying in the range 15--24\,$\mu$Jy, that is the $3\sigma$ detection threshold estimated by the map noises listed in Table~\ref{tab:vla_obs}) to constrain the mass-loss rate of the stellar wind.

In the absence of a strong globally organized magnetic field, the mass-loss rate, $\dot{M}(B=0)$, could be theoretically predicted and empirically measured by conventional spectroscopic methods. The strong, likely dipolar, magnetic field of the qWR star should strongly alter the topology of its stellar wind. While at low magnetic latitudes the wind is trapped, it is free to escape from the magnetic polar regions. It is reasonable to expect that the radial component of the polar wind is larger compared to the wind at low magnetic latitudes, which is trapped by the magnetic field and forced to flow along the magnetic field lines, thus the wind speed has no radial component near the magnetic equator. This may explain the differences between the wind terminal velocities at the equator and the pole ($V_{\infty}^{\rm eq} \approx 425$\,km\,s$^{-1}$ and $V_{\infty}^{\rm pole}\approx 1200$\,km\,s$^{-1}$; \citealp{Groh2008}).

The wind of the qWR star is expected to be confined by the dipole magnetic field up to the Alfv\'en radius, and only beyond it the ionized material can escape like a nearly spherical non-magnetic wind. At a large distance, the stellar wind which emerges from the polar regions fills a major fraction of the spherical volume. Therefore, the upper limit on radio emission allows to establish an upper limit on the mass-loss rate of the escaping wind. The strongest constraint for the mass loss rate of the freely escaping wind is given by the observation in the 15\,GHz band which has the highest sensitivity (see Table~\ref{tab:vla_obs}) corresponding to a $3\sigma$ detection threshold of 15\,$\mu$Jy. We calculated radio spectra using the Eq.\,(\ref{eq:windspec1}) varying only the parameter $\dot{M}$, while the terminal velocity was fixed to $V_{\infty}^{\rm pole}=1200$\,km\,s$^{-1}$.

\begin{figure}[]
\centering
\includegraphics[width=1.\columnwidth]{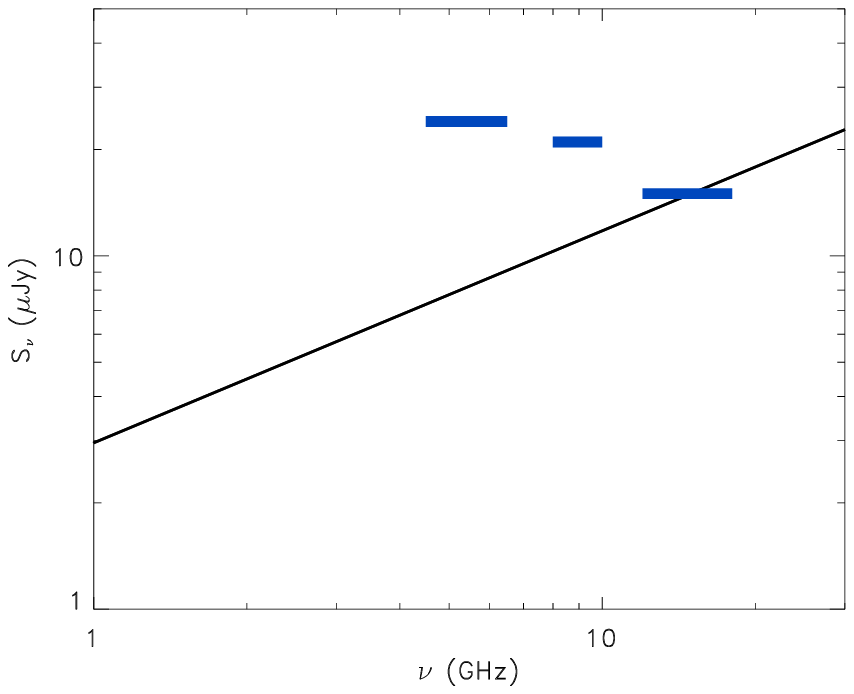}
\vspace{-2mm}\\ 
\caption{
The blue boxes correspond to the $3\sigma$ noise levels measured at the sky position of \qwr\ in radio maps obtained at the three observed bands, providing the detection threshold. The lengths of blue boxes show the bandpass for the receivers (see Table~\ref{tab:vla_obs}). The black solid line displays the theoretical thermal wind's spectrum corresponding to the maximum emission level that is compatible with the non-detection below the $3\sigma$ threshold ($\dot{M}_{\mathrm{max}}=1.1 \times10^{-7}$\,M$_{\odot}$\,yr$^{-1}$).
}
\label{fig:radio_only_wind}
\end{figure}

The maximum mass-loss rate of the qWR star producing thermal radio emission with an emission level that does not contradict the VLA non-detection is $\dot{M}_{\mathrm{max}}=1.1 \times10^{-7}$\,M$_{\odot}$\,yr$^{-1}$, which is about a factor of two lower than the mass-loss rate of $2.2 \times 10^{-7}$ M$_{\odot}$\,yr$^{-1}$ found by \citet{Groh2008}. The corresponding theoretical wind spectrum is pictured in Fig.~\ref{fig:radio_only_wind} (black solid line). Assuming as a check the slower equatorial wind velocity, $V_{\infty}^{\rm eq}=425$\,km\,s$^{-1}$, a mass-loss rate of about $3.9\times10^{-8}$ M$_{\odot}$\,yr$^{-1}$, that is lower than $\dot{M}_{\mathrm{max}}$, is required to be consistent with the upper limits in radio. Therefore, our radio observations definitively rule out the wind parameter combination corresponding to the lower value of the terminal wind velocity, $V_{\infty}=425$\,km\,s$^{-1}$, and the mass-loss rate of $2.2 \times 10^{-7}$ M$_{\odot}$\,yr$^{-1}$ \citep{Groh2008} as global properties of the wind escaping from the qWR star.

\section{X-rays from the wind of the highly magnetized helium star in the \qwr\ system}
\label{sec:wind_and_xadm}

 \begin{figure*}[]
\centering
\includegraphics[width=1.8\columnwidth]{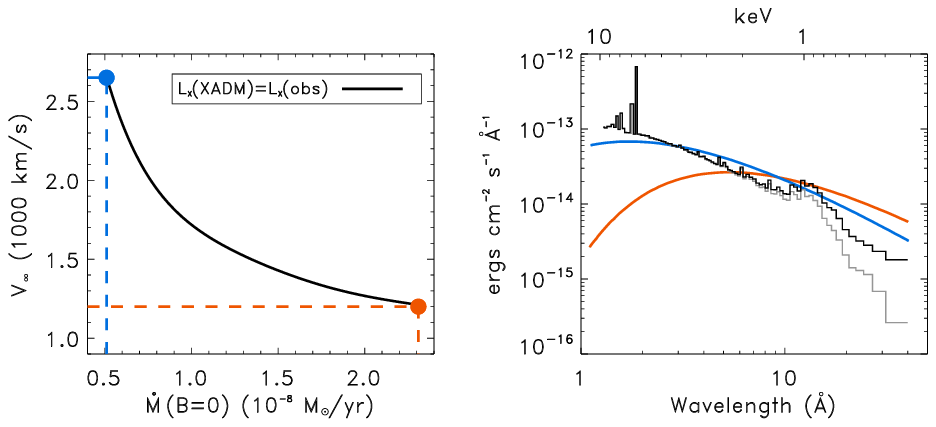}
\caption{
Left: Diagram of the wind parameters of a non-magnetized star; terminal velocity vs mass-loss rate (the two wind parameters are the free parameters of the XADM model). The black curve represents the loci of the $\mdot(B=0)$ and $V_{\infty}$ combination where the XADM model predicts X-ray luminosity coinciding with the empirically determined for the qWR star (Table~\ref{tab:spec}). The red and blue horizontal dashed lines mark respectively the lower and upper limits of the range of terminal wind speed values analyzed, with $\mdot(B=0)$ tuned to match the observed X-ray luminosity along the black contour ($\log (L_{\mathrm X} / L_{\mathrm {bol}}) = -5.6$). Right: Associated model X-ray spectra for the fast (blue solid line) and the slower wind model (red solid line). For comparison, the X-ray spectrum of \qwr\ provided by the 3T model without the ISM absorption effect is also displayed (black solid line), the grey solid line is the X-ray spectrum with absorption shown in the left panel of Fig.~\ref{fig:spec-uf}.
}
\label{fig:XADMfigs}
\end{figure*}

\subsection{Thermal X-ray emission from the magnetic qWR} 
\label{sec:xraytermo}

To quantify the X-rays emitted by the plasma heated in the framework of the MCWS model, time-dependent MHD simulations included a full energy equation with radiative cooling were performed by \cite{ud-doula2014}; their analysis of how the associated X-ray emission is controlled by a cooling-regulated `shock retreat' led to a semi-analytic `XADM' formalism. Once assigned the luminosity, mass, and radius of the star, the XADM model allows to predict the level and hardness of the intrinsic emitted X-rays as a function of the polar strength of the dipole field $B_{\mathrm p}$, the mass-loss rate ${\dot M(B=0)}$, and the terminal speed $V_\infty$ of the stellar wind  expected in the absence of magnetic field. The agreement with observations is reached by scaling the predicted X-ray luminosity by an order of magnitude \citep{Naze2014}. The 10\% empirical reduction of the XADM prediction is to account for the lower X-ray emission expected from the dynamic infall of the trapped material, which is not considered in the idealized XADM model. In the following, we scale down the XADM predictions by an order of magnitude.

We modify the XADM formalism to account for highly unusual properties of the qWR star in \qwr. For simplicity, we assume a pure helium wind. Furthermore, the polar field strengths is assumed equal to the average field ($\langle B \rangle=43$\,kG). Usually, the polar strength is expected higher than the average, but the particular geometry of the qWR (i.e. the dipole is seen pole-on) makes this approximation reasonable.

Using stellar radius and bolometric luminosity reported in Table~\ref{tab:qwr}, the left panel of Fig.\,\ref{fig:XADMfigs} shows the contour of the predicted $\log (L_{\mathrm X} / L_{\mathrm {bol}})$ (above 0.3 keV) set to the luminosity derived from the X-ray observations ($\log (L_{\mathrm {X,\,obs}} / L_{\mathrm {bol}}) = -5.6$; Table~\ref{tab:spec}) as a function of non-magnetized wind mass-loss rate and terminal speed. The horizontal blue and red dashed lines mark respectively the limits of the range of the terminal speed values explored. For convenience the lower limit ($V_{\infty}^{\mathrm{slow}}$) has been fixed at 1200\,km\,s$^{-1}$, that is the empirical wind speed value \citep{Willis1989}, and the upper limit ($V_{\infty}^{\mathrm{fast}}$) at 2650\,km\,s$^{-1}$, which is the value predicted by \citet{Vink2017} for a 2\,M$_{\odot}$ helium star without magnetic field. The corresponding vertical dashed lines mark the associated mass-loss rates, $\dot M(B=0) =0.51 \times 10^{-8}$\,\myr\ and $2.3 \times 10^{-8}$\,\myr.

The right panel of Fig.\,\ref{fig:XADMfigs} compares the model  X-ray spectra corresponding to the velocities extrema, slow (red) and fast (blue). The XADM model with the fast wind has a hard X-ray component not predicted by the model with slower wind, with also a steeper overall decline at longer wavelengths. The qualitative comparison between the XADM synthetic spectra with the best-fit X-ray spectrum, provided by the pure thermal model once the ISM absorption effect is removed, evidences that the X-ray spectral behavior of \qwr\ is closer to the XADM 
{fast wind model, giving also the feeling that the terminal wind speed could be higher than the fixed upper limit.}

Thus the XADM modeling provides an estimate of the combination of wind parameters (mass-loss rate and terminal wind speed) by comparing the shapes of the simulated and observed X-ray spectra. Our analysis showed that a wind with parameters close to those theoretically predicted for a non-magnetic naked helium star with 2 solar masses (${\dot M_{\rm{th}}} \approx 0.4 \times 10^{-8}$\,\myr and $V_{\infty}^{\rm th}  = 2650$\,km\,s$^{-1}$; \citealp{Vink2017}) is consistent with the X-ray observations.

\begin{table*}[!htb]
\centering
\caption{Magnetospheric parameters of the qWR star. }
\begin{tabular}{lccc}
\hline  \hline
Parameter &Symbol & Unit & Value  \\
\hline
\vspace{-3mm}\\ 

Mass-loss rate in absence of magnetic field derived from the XADM model   & $\dot{M}(B=0)$ & \myr\ &  {$5.1\times10^{-9}$} \\
Terminal wind velocity theoretically predicted in absence of magnetic field    & $ {V_{\infty}^{\rm th}}^{\dag} \equiv {V_{\infty}^{\rm fast}}$ & km\,s$^{-1}$            & 2650 \\
Wind confinement parameter  & $\eta_{\ast}$ & -  & {20500} \\
Alfv\'en radius &$R_{\mathrm {A}}$ & R$_{\ast}$  & {12.3} \\
Mass-loss rate actually lost from the magnetic qWR & $\dot{M}$ & \myr\ &  {$\approx 3\times10^{-10}$} \\
Radial wind velocity measured from spectroscopy             & $V_\infty({\mathrm{obs.}}) ^{\dag\dag} \equiv {V_{\infty}^{\rm slow}}$ & km\,s$^{-1}$            & 1200 \\
Kepler radius & $R_{\mathrm K}$ & R$_{\ast}$  & $\approx 150$ \\
\hline

\end{tabular}

$^{\dag}$  \cite{Vink2017}.
$^{\dag\dag}$ \cite{Willis1989};

\label{tab:deriv_param}
\end{table*}

\subsection{{Constraints} on the wind's parameters  of the magnetic qWR}
\label{sec:wind_and_magfield}

In plasma theory, the ratio between the gas pressure and the magnetic pressure of a stationary plasma defines the parameter $\beta$. In the case of a supersonic stellar wind channeled by a dipolar magnetic field, the thermal pressure is replaced by the ram pressure of the wind \citep{Altschuler1969}. Then, the reciprocal of plasma-$\beta$ is:  
\begin{equation}
\eta_{\ast}=\frac{B^2/8\pi}{\rho_{\mathrm w} V_{\infty}^2 /2} \approx \frac{B_{\mathrm p}^2 R_{\ast}^2}{4 \mdot(B=0) V_{\infty}},
\label{eq:eta}
\end{equation}
that is the magnetic confinement parameter which characterizes the capability of the stellar magnetic field to channel the wind, which starts on the stellar surface as a radial wind, that is, as a simple non-magnetic spherical wind \citep{ud-doula2002}. In this description, the Alfv\'en radius is given by the following relation \citep{ud-doula2008}:
\begin{equation}
R_{\mathrm {A}} \approx 0.3 + \eta_{\ast}^{1/4}.
\label{eq:ralf}
\end{equation}

Assuming a simple dipole magnetic field, the wind emerging from the stellar surface ($\mdot(B=0)$) could actually leave the magnetosphere only from the magnetic polar caps. The polar caps are delimited by the northern and southern polar rings located by the magnetic latitude of the footprints of the last closed magnetic field line, that is the line crossing the magnetic equatorial plane at the distance equal to $R_{\mathrm {c}} \approx 1 + 0.7( R_{\mathrm {A}} -1)$ stellar radii. Following \citet{ud-doula2008}, the wind material lost from a magnetic star can be estimated using the scaling relation: 
\begin{equation}
\mdot   \approx \mdot(B=0) \times (1 - \sqrt {1 - 1 / R_{\mathrm {c}}}).
\label{eq:mdotact}
\end{equation}

With a polar magnetic field strength equal to 43\,kG, the associated magnetic confinement parameter, calculated using the Eq.\,(\ref{eq:eta}) ($V_\infty = 2650$\,km\,s$^{-1}$ and $\dot M(B=0) =0.51 \times 10^{-8}$\,\myr), is {$\eta_\ast=20500$. The corresponding Alfv\'en radius calculated using Eq.~(\ref{eq:ralf}) is $R_{\mathrm A}/R_\ast = 12.3$. 

In summary, the wind of the qWR star is expected to be confined by the closed dipole magnetic field lines up to the Alfv\'en radius, and only beyond this radius the ionized material can escape like a nearly spherical non-magnetic wind. The fraction of the wind mass that escapes can be estimated using the Eq.~(\ref{eq:mdotact}), that is about 6\% of the wind emerging from the whole stellar surface. The corresponding actual mass-loss rate is $\approx 3 \times 10^{-10}$\,\myr, that is about three orders of magnitudes lower than the upper limit constrained by the radio observations (Sect.\,\ref{sec:radio_diagn}), with a related thermal radio emission level of $\approx 5\times10^{-3}$\,$\mu$Jy (calculated using Eq.\,(\ref{eq:windspec1})).

The wind topology of the qWR is strongly affected by the strong magnetic field. Only a fraction of the fast wind (${V_{\infty}^{\rm fast}}=2650$\,km\,s$^{-1}$) can escape from the magnetosphere at high magnetic latitudes, becoming nearly spherical far from the star. Due to the larger area covered by the freely expanding spherical wind, according to the principle of mass continuity,  the non-magnetic spherical wind is expected to expand outside the Alfv\'en surface at a slower velocity with respect to the fast wind velocity required to reproduce the X-ray spectrum of the qWR star. Therefore, the velocity measured in the UV spectrum ($V_\infty({\mathrm{obs.}})=1200$\,km\,s$^{-1}$) is considered as an empirical estimate of the terminal wind velocity of the large-scale spherical wind material actually lost from the qWR star. Figure\,\ref{fig:scenario} shows a cartoon that visualizes the scenario described above.

\section{On the lack of detection of non-thermal radio emission from the qWR in \qwr}
\label{sec:nonterm_radio}

The qWR star is strongly magnetized, and the combined presence of ionized material (wind plasma trapped by the closed magnetic field lines) and strong magnetic field could provide suitable conditions for non-thermal radio emission powered by the gyro-synchrotron mechanism. In the following we use the detection threshold obtained by the VLA observations to constrain the plasma parameters regulating a non-thermal radio emission that is compatible with the non-detection of radio emission from \qwr.

\subsection{On the non-thermal electron production within the magnetosphere of the qWR}
\label{sec:accel_mechanism}

To better understand the properties of the qWR star, we first consider the case of magnetic ApBp-stars. These are also hot stars with radiatively driven stellar winds confined by strong magnetic field. Hence, ApBp-stars provide a useful physical analogy to the ``one of its kind'' magnetic helium star in \qwr. The magnetic ApBp-stars have magnetospheres filled by plasma radiating X-rays; furthermore, these magnetic stars are well-studied non-thermal radio sources. The well-ordered and stable magnetospheres of fast-rotating ApBp stars share a common physical mechanism for supporting the generation of non-thermal electrons, which explains their observational features ranging from the X-ray to radio regime. A general scaling relationship for non-thermal radio emission holds from stars at the top of the main sequence down to the ultra-cool dwarfs and to the planet Jupiter \citep{Leto2021}. This is empirically confirmed for ApBp stars \citep{Shultz2022}. The underlying physical process could be related to the breakout events that are predicted in the centrifugally supported magnetospheres (CM) of fast rotating stars \citep{Owocki2022}. In this case, the radio luminosity is directly related to the power released by the centrifugal breakout events (CBO). In particular, the relation between the spectral radio luminosity and the power of the centrifugal breakouts ($L_{\mathrm {CBO}}$) is: $L_{\nu,{\mathrm{rad}}}=10^{-19}L_{\mathrm {CBO}}$\,Hz$^{-1}$ \citep{Leto2022}, where
\begin{equation}
\label{eq:lumcbo1}
L_{\mathrm{CBO}}=\frac{{B_{\rm p}^2R_{\ast}^3}}{P_{\rm rot}}\times W~~~{\mathrm{(erg \,s^{-1})}},
\end{equation}
\noindent 
and $W=2 \pi R_{\ast} / (P_{\rm rot}\ \sqrt{GM_{\ast}/R_{\ast}})$ is the dimensionless critical rotation parameter, defined by the ratio between equatorial velocity and orbital velocity, with $G$ being the gravitational constant. 

Using the stellar parameters reported in Table~\ref{tab:qwr} it follows that $W\approx5.4\times10^{-4}$ for the qWR star. Eq.~(\ref{eq:lumcbo1}) predicts the spectral radio luminosity $L_{\nu,{\mathrm{rad}}}\approx2.1 \times 10^{12}$\,erg\,s$^{-1}$\,Hz$^{-1}$, corresponding to the flux of $\approx 1.8\times 10^{-3}$\,$\mu$Jy, which, in the frequency range covered by the VLA observations, is lower than the expected flux level of the thermal radio emission constrained in Sect.\,\ref{sec:wind_and_xadm}. Hence, the CBOs are inefficient as a mechanism for the acceleration of electrons in the magnetic qWR star. This is not surprising since the key condition for the generation of CMs is the Kepler corotation radius ($R_{\mathrm K}$) which is smaller than the Alfv\'en radius. The Kepler radius is given by $R_{\mathrm K}=W^{-2/3} R_{\ast}$ \citep{Petit2013}. We estimate $R_{\mathrm K} \approx 150$\,R$_{\ast}$, which is about an order of magnitude larger than the Alfv\'en radius estimated in Sect.~\ref{sec:wind_and_magfield} (Table\,\ref{tab:deriv_param}), this implies that the qWR does not support a centrifugal magnetosphere.

Let's consider whether stellar wind might play a role in particle acceleration. Indeed, the wind strength in the qWR star is much higher compared to the ApBp-type stars \citep{Osk2011, Krt2019}, and can break the magnetic field lines relatively close to the star ($R_{\mathrm A}\approx 12$\,R$_{\ast}$) where the local magnetic field is high enough to trigger plasma effects responsible for non-thermal radio emission \citep{Andre1988}. At distances larger than $R_{\mathrm A}$, the magnetic fields no longer controls the wind. The equatorial regions just outside the Alfv\'en surface are transition regions from the magnetic-dominated to the wind-dominated zones, where the closed dipole-like magnetic field configuration becomes open. These are likely sites of large-scale magnetic field reorganization, with the consequent formation of a magneto-disk, possibly originating current sheets. In these regions, magnetic field lines of opposite polarity exist and magnetic reconnections may occur, where electrons can be accelerated up to relativistic energies \citep{Usov_Melrose1992}. This acceleration mechanism has been also taken into account to predict the emission of gamma rays from a magnetic hot massive star with a strong wind and DM magnetosphere \citep{Bednarek2021}.

\subsection{Modeling the non-thermal radio emission from the qWR}
\label{sec:nontermspec}

Both ingredients for the production of non-thermal radio emission -- magnetic field and relativistic electrons -- could be expected in the magnetosphere of the qWR star. The predicted level and the spectral shape of emerging non-thermal radio radiation depend on two key factors: (1) the acceleration efficiency which dictates how much non-thermal radio radiation is produced, and (2) the wind optical depth which determines how much non-thermal radiation is able to escape.

To calculate the spectrum of non-thermal radio emission we employ 3D model of a dipole shaped magnetosphere \citep{Trigilio_etal2004,Leto2006}. After sampling the space surrounding the star by using a cartesian grid with three different sampling steps, once assigned the radio frequency ($\nu$), all the physical parameters needed for the calculation of the gyro-synchrotron absorption and emission coefficients are calculated in each grid point and the radiative transfer equation is numerically integrated along ray paths parallel to the line of sight.The orientation of the dipole-shaped magnetosphere with respect to the observer can be arbitrarily varied, this allows a proper reproduction of the measured rotational modulation of the non-thermal radio emission from co-rotating magnetospheres of ApBp-type stars with magnetic dipole axis not aligned to the rotation axis \citep{Leto2012,Leto2017,Leto2018,Leto2020,Leto2020b}. Even if the most likely magnetic field geometry of the qWR star suggest a pole-on view \citep{Shenar2023}, for completeness, we also performed model simulations for the equator-on view (corresponding to the magnetic axis perpendicular to the line of sight).

To compute the gyro-synchrotron radio spectrum of the qWR, we conservatively assumed a polar field strength $B_{\mathrm p}=43$\,kG, as already discussed in Sect.\,\ref{sec:xraytermo}. The relativistic electrons are assumed power-law energy distributed, $N_{\mathrm{rel}}(E) \propto E^{-\delta}$, where $N_{\mathrm {rel}}(E)$ is the number density of the electrons of energy $E$. Similar to the case of the ApBp-magnetic stars \citep{Leto2021}, a low energy cutoff at $E=10$\,keV for the non-thermal electrons is adopted, and the spectral index of the energy distribution of the non-thermal electrons is assumed as $\delta=2.5$. The region where these non-thermal electrons move is defined by the Alfv\'en radius. We assume  that the  thermal plasma density located below $R_{\mathrm {A}}$ is a function of both the radial distance and the magnetic colatitude. For crude estimates, we make use of the density law valid for a magnetically trapped wind retrieved by modeling a slow-rotating dynamical magnetosphere with an analytic approach (ADM model) \citep{Owocki2016}. The adopted parameters of the trapped wind are $\dot{M}(B=0)$ and $V_{\infty}^{\rm th}$, both listed in Table\,\ref{tab:deriv_param}.

\begin{figure}[]
\centering
\includegraphics[width=1.\columnwidth]{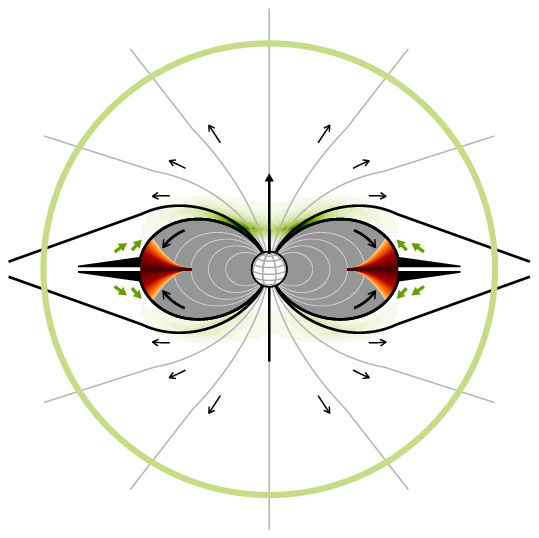}
\vspace{-2mm}\\ 
\caption{
Meridional cross-section of the dipole-dominated magnetosphere of \qwr\ (not to scale). At distances smaller than the Alfv\'en radius ($R_{\mathrm A}$) the magnetic field lines are closed. The last closed field lines (marked by the black thick solid line) locates the $R_{\mathrm A}$. At larger distances, the ionized wind opens the magnetic field lines. The region where the magnetic field traps the stellar wind is shaded. Outside of the Alfv\'en surface, the stellar wind freely escapes (black arrows). At lower latitudes, the fast wind plasma streams arising from opposite hemispheres (thick black arrows) collide and shock. According to the XADM model, the shock heats the plasma producing X-rays (red regions). The equatorial magneto-disk farther than $R_{\mathrm A}$ (thick black areas) is likely the site of acceleration of electrons (green solid arrows) via magnetic reconnections. Such non-thermal electron population moving within a magnetic shell (delimited by the black and thick solid lines) radiates at the radio regime by the gyro-synchrotron emission mechanism. The outer boundary where the non-thermal electrons diffuse is pictured by the solid black open field line. The relativistic electrons radiate in radio bands by a non-thermal emission mechanism (green-shaded regions). The large green circle outlines the optically thick radio photosphere of the freely escaping ionized wind.
} 
\label{fig:scenario}
\end{figure}

The only free parameter is the column density of the relativistic electrons injected at the distance of $R_{\mathrm A}$, which is the number density of the relativistic electron times the equatorial linear size of the magnetic shell where they freely move. This size is related to the size of the magneto-disk where the non-thermal electrons are likely accelerated. Finally, to calculate the emerging non-thermal radio spectrum of the qWR, we also account for the attenuation of radio waves traveling through the outer layers filled by thermal plasma coming from the stellar wind, which can be parameterized by the radius of the wind region that is optically thick at a given radio frequency (see Figure\,\ref{fig:scenario}).

According to \citet{Panagia1975}, the radius of the optically thick radio photosphere of a spherical ionized wind is: 
  \begin{align}
R_{\nu} \approx 6.23 \times 10^{14}
\left(\frac{\nu}{{10\,\mathrm{GHz}}}\right)^{-0.7}  
\left(\frac{T_{\mathrm{e}}}{10^4 \,{\mathrm K}}\right)^{-0.45} \nonumber  
\left(\frac{\dot M}{10^{-5} \,{\mathrm {M_{\odot} \, yr^{-1}}}}\right)^{2/3} \\
\left(\frac{V_{\infty}}{10^{3} \,{\mathrm {km \, s^{-1}}}}\right)^{-2/3}\,
\mathrm{cm.}
  \label{eq:radius_radio_phot}
  \end{align}

\noindent
Using the mass-loss rate of the wind material actually lost from the magnetic qWR star ($\dot{M} = 3\times10^{-10}$\,M$_{\odot}$\,yr$^{-1}$) and assuming spherically expanding wind outside the Alfv\'en radius ($R_{\mathrm A}\approx 12$\,R$_{\odot}$); see Sect.\,\ref{sec:wind_and_magfield}) with velocity $V_{\infty}(\mathrm{obs.})=1200$\,km\,s$^{-1}$, the radius of the radio photosphere at $\nu=100$\,MHz is $R_{0.1\,{\mathrm{GHz}}} \approx 110$\,R$_{\ast}$, that at $\nu=10$\,GHz decreases to $R_{10\,{\mathrm{GHz}}} \approx 4$\,R$_{\ast}$ that is lower than $R_{\mathrm A}$.

To search for suitable conditions for the VLA non-detection, we computed models progressively decreasing the column density of the non-thermal electrons. We performed model simulation covering a wide frequency range extending to the low frequencies domain. Outside the magnetospheric volume, the wind material freely escaping has been roughly assumed spherical, and the continuity equation $\rho=\dot{M}/4\pi r^2 v(r)$ of the wind expanding according to a velocity law $v(r)=V_{\infty}(1 - R_{\ast}/r)$ was adopted. To include absorption, we integrate the radiative transfer equation within a huge cubic volume with lateral size much larger than the radio photosphere corresponding to the lowest analyzed frequency (100\,MHz), i.e.\ 600\,R$_{\ast}$. The small cubic element with a side of 0.25\,R$_{\ast}$ is used to sample the inner cube with side 15\,R$_{\ast}$. The intermediate sampling of 0.5\,R$_{\ast}$ was adopted for the regions within the cube with side 40\,R$_{\ast}$. At larger distances a sampling step of 4\,R$_{\ast}$ is adopted.

The simulated gyro-synchrotron spectra computed for the two extrema magnetospheric orientations (pole-on view and equator-on view) and} 
compatible with the VLA upper limits are pictured in Fig.~\ref{fig:radiodotm1} (grey region). The corresponding limit on the non-thermal electrons column density is $\approx 3 \times 10^{15}$\,cm$^{-2}$. For comparison, in fast rotating early-type magnetic stars, the column density of non-thermal electrons able to reproduce their observed gyro-synchrotron radio spectra is on average $10^{16}$\,cm$^{-2}$ \citep{Leto2021}. This evidences that, even assuming a radio emission level just below the VLA detection threshold, the possible acceleration mechanisms of non-thermal electrons operating within the DM of the qWR star in \qwr\ is likely less efficient than the CBOs inside the centrifugal magnetospheres surrounding the ApBp-type stars.

The wind's radio photosphere depends on frequency, as $\propto \nu^{-0.7}$  (Eq.~(\ref{eq:radius_radio_phot})), therefore, the wind absorption effect increases at lower frequencies. In fact, looking at Fig.\,\ref{fig:radiodotm1}, the possible non-thermal gyro-synchrotron emission from the qWR star is fully absorbed at the lower frequency range. Despite the rough sampling step adopted at a large distance, this discrete integration of the radiative transfer equation produces a fully absorbed non-thermal radio spectrum at the lower frequencies ($\nu \lessapprox 300$\,MHz) almost indistinguishable from the theoretical radio spectrum of a spherical wind predicted by the Eq.~(\ref{eq:windspec1}) (adopted wind parameters: $\dot{M}=3\times10^{-10}$\,M$_{\odot}$\,yr$^{-1}$; $V_{\infty}=1200$\,km\,s$^{-1}$) depicted in Fig.~\ref{fig:radiodotm1} by the black dashed line, with the wind absorption effect that is non negligible for the gyro-synchrotron emission up to frequencies close to $\approx 1$\,GHz. On the other hand, the wind actually lost from the qWR is too weak to provide significant absorbing effects in the VLA observing bands.

\subsection{Other possible cases for the model application}
\label{sec:discussion}

\begin{figure}[]
\centering
\includegraphics[width=1.\columnwidth]{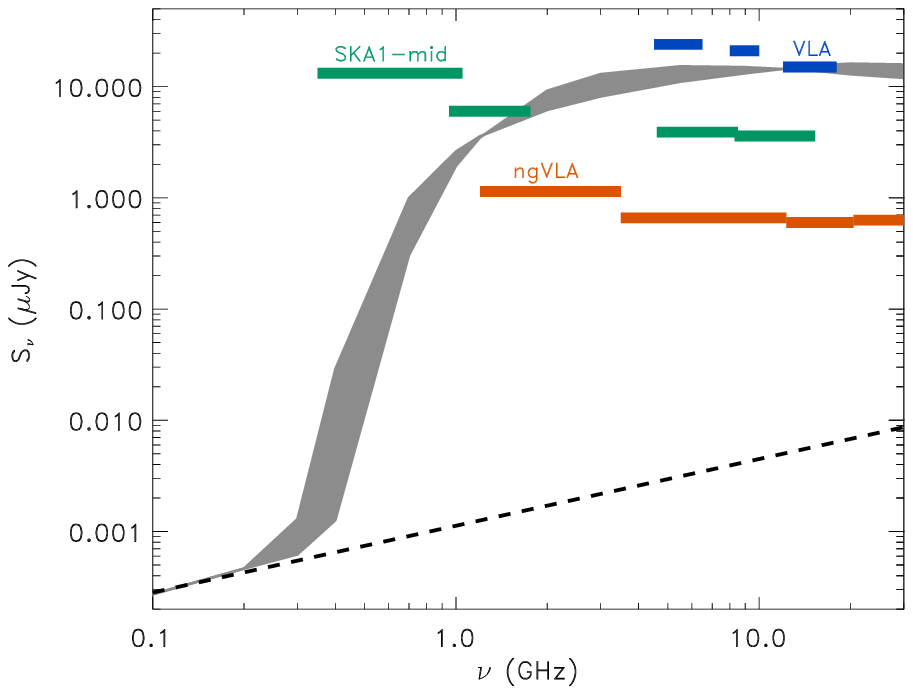}
\vspace{-2mm}\\ 
\caption{
Theoretical radio spectra of the qWR component in the \qwr\ system in comparison with the upper limits derived from the VLA observations (blue boxes). The black dashed line is for a spherical wind with mass-loss rate $\dot{M} = 3  \times 10^{-10}$\,M$_{\odot}$\,yr$^{-1}$ and $V_{\infty}=1200$\,km\,s$^{-1}$. The grey shaded area shows the variety of non-thermal gyro-synchrotron spectra computed for different orientations of the magnetosphere with respect to the observer, assuming two distinct values of the spectral index $\delta$, and accounting for the attenuation in the stellar wind. The green boxes represent the expected $3\sigma$ threshold of the noise levels that will be reached with observations 1 hour long at the observing bands that will be provided by the forthcoming SKA1-mid. The orange boxes instead represent the expected $3\sigma$ threshold of the noise levels that will be achieved assuming the same integration time (1 hour) using the observing bands that will be provided by the future ngVLA.
}
\label{fig:radiodotm1}
\end{figure}

The modeling approach presented in this paper represents the computational implementation of the qualitative model of a non-thermal stellar radio source embedded within a large-scale spherical environment of ionized material released by the freely escaping wind \citep{Andre1988} which should be applied also in cases of other hot magnetic stars that originate powerful stellar winds. I.e. the highly magnetized O7-type star NGC\,1624-2, which is intrinsically very bright at the X-rays, but its X-ray emission is strongly attenuated by the wind material trapped within the stellar magnetosphere \citep{Petit2015} and, similarly to \qwr, NGC\,1624-2  was undetected at the radio regime \citep{Kurapati2017}. Or to the case of the magnetic O7.5-type star member of the O+O binary system HD\,47129, which has similar X-ray luminosity of \qwr\ \citep{Naze2014} and further shows an interesting radio spectral behavior. This star was undetected at the lower frequency but has an almost flat emission at the other two highest frequencies \citep{Kurapati2017}. The spectral range covered by the HD\,47129 radio measurements covers the spectral range of the radio measurements here reported. We highlight that the observed spectral behavior of HD\,47129 is qualitatively in accordance with the synthetic radio spectra reported in this paper (see Fig.\,\ref{fig:radiodotm1}), but shifted in frequency, supporting the idea that non-thermal gyro-synchrotron radio emission may undergo frequency-dependent absorption effects provided by the large-scale surrounding ionized medium.

\section{Summary and Conclusions}
\label{sec:conclusions}

In this paper we report \xmm\ and VLA measurements of the \qwr\ system composed by a non magnetic late B-type star and a highly magnetized quasi-WR star. The system has been clearly detected at the X-rays, but no radio counterpart has been found by the highly sensitive radio measurements.

The upper limit on radio emission allowed us to put robust observational constraints on the actual wind mass-loss rate of the the qWR star. Additionally, the detection of the emission line at $\lambda=1.87$\,\AA\ ($\sim 6.6$\,keV) in the X-ray spectrum is the observational evidence that very fast plasma streams exist within the magnetosphere of the qWR. The XADM model is able to explain the observed X-ray luminosity and spectrum shape (see Fig.\,\ref{fig:XADMfigs}). On the other hand, the magnetic field traps a large fraction of the wind reducing the amount of material effectively lost. We constrain the effective wind mass-loss rate of the magnetic qWR $\dot{M} \approx 3\times 10^{-10}$\,\myr. 

The qWR star component of the \qwr\ system has a mass of $2.03\pm0.44$\,M$_{\odot}$, which is at least 0.15 solar masses above the Chandrasekhar limit of 1.44\,M$_{\odot}$. The time required to remove enough material via stellar wind to fall below the Chandrasekhar limit is $500$\,Myr, i.e. longer than the helium burning phase of a massive star with a He core of similar mass to the qWR which last a few Myr \citep{Ritter2018}.

We also explored the physical mechanisms able to produce radio emission from the magnetic qWR star, then providing constraints on the parameters that make it compatible with the upper limits measured by the VLA. We calculated the non-thermal radio spectrum taking into account the extinction effects. To explain that there is no non-thermal radio emission above the VLA detection threshold, we found that the non-thermal electron production has to be less efficient compared to the fast-rotating ApBp-type magnetic stars. 
{Generally, in ApBp-type magnetic stars the gyro-synchrotron radio emission level correlates with the stellar rotation speed \citep{Leto2021} and therefore to the centrifugal breakouts power, which is the proposed driving mechanism for non-thermal electron production \citep{Owocki2022}.}
This is not the case of  the magnetic qWR star in \qwr\ because it is a slow rotator with a DM, therefore not able to efficiently sustain rotationally supported acceleration mechanisms. Hence the non-thermal electron production is fundamentally different and less efficient compared to ApBp-type stars with CMs. 

Studying of the radio spectrum of \qwr\ is an ideal science case for the SKA1-mid radio interferometer. The expected noiselevel at the available radio bands of SKA1-mid, operating between $350$\,MHz and 15\,GHz, ranges between $\approx 1$ and $\approx 4$ $\mu$Jy for observations 1 hour long (green boxes in Fig.~\ref{fig:radiodotm1}) \citep{Braun2019}. But even more powerful will be the future next generation VLA (ngVLA). Assuming radio observations 1 hour long, the ngVLA will allow us to explore the radio spectrum of \qwr\ with a sub $\mu$Jy sensitivity level (orange boxes of Fig.~\ref{fig:radiodotm1}) at all the available frequency bands (which will cover the frequency range from 1.2 up to 116 GHz) \citep{Selina2018}. The ngVLA will be able to definitely test the presence of non-thermal radio emission in \qwr\ to unveil the physical processes occurring within the magnetosphere of the qWR, which is a DM prototype.

To conclude, we have empirically demonstrated that the acceleration processes (if any) occurring within the magnetosphere of the slowly rotating qWR magnetic star in the \qwr\ system are less efficient than that occurring within the CMs of the fast rotating ApBp-type stars. We also evidenced that the theoretical recipe to estimate the wind parameters from non-magnetic helium stars fits well the case of the magnetically constrained wind of the qWR. Further, taking into account the magnetic nature of this evolved star, we estimated that the actual rate of mass lost from the magnetic qWR star is not able to remove a significant amount of mass. Hence, this (currently) unique object is likely to maintain its super-Chandrasekhar mass until its death and undergo a core-collapse supernova explosion producing a magnetar.

\begin{acknowledgements}
We thank the referee for his/her constructive comments that helped us to improve the paper.
SO and AuD acknowledge support by the National Aeronautics and Space Administration under Grant No. 80NSSC22K0628 issued through the Astrophysics Theory Program.
RI gratefully acknowledges support by the National Science Foundation under grant number AST-2009412.
GAW acknowledges Discover Grant support from the Natural Sciences and Engineering Research Council (NSERC) of Canada.
\end{acknowledgements}

%
%

\bibliographystyle{aa} 
\bibliography{papers,mag}

\begin{thebibliography}{74}
\expandafter\ifx\csname natexlab\endcsname\relax\def\natexlab#1{#1}\fi

\bibitem[{{Altschuler} \& {Newkirk}(1969)}]{Altschuler1969}
{Altschuler}, M.~D. \& {Newkirk}, G. 1969, Solar Physics, 9, 131

\bibitem[{{Andre} {et~al.}(1988){Andre}, {Montmerle}, {Feigelson}, {Stine}, \&
  {Klein}}]{Andre1988}
{Andre}, P., {Montmerle}, T., {Feigelson}, E.~D., {Stine}, P.~C., \& {Klein},
  K.-L. 1988, \apj, 335, 940

\bibitem[{{Anger}(1933)}]{Anger1933}
{Anger}, C.~J. 1933, Harvard College Observatory Bulletin, 891, 8

\bibitem[{{Arnaud}(1996)}]{xspec}
{Arnaud}, K.~A. 1996, in Astronomical Society of the Pacific Conference Series,
  Vol. 101, Astronomical Data Analysis Software and Systems V, ed. G.~H.
  {Jacoby} \& J.~{Barnes}, 17

\bibitem[{{Babcock}(1960)}]{Babcock1960}
{Babcock}, H.~W. 1960, \apj, 132, 521

\bibitem[{{Babel} \& {Montmerle}(1997)}]{Babel1997}
{Babel}, J. \& {Montmerle}, T. 1997, \aap, 323, 121

\bibitem[{{Bailer-Jones} {et~al.}(2021){Bailer-Jones}, {Rybizki}, {Fouesneau},
  {Demleitner}, \& {Andrae}}]{Bailer-Jones2021}
{Bailer-Jones}, C.~A.~L., {Rybizki}, J., {Fouesneau}, M., {Demleitner}, M., \&
  {Andrae}, R. 2021, \aj, 161, 147

\bibitem[{{Bednarek}(2021)}]{Bednarek2021}
{Bednarek}, W. 2021, \mnras, 507, 3292

\bibitem[{{Braun} {et~al.}(2019){Braun}, {Bonaldi}, {Bourke}, {Keane}, \&
  {Wagg}}]{Braun2019}
{Braun}, R., {Bonaldi}, A., {Bourke}, T., {Keane}, E., \& {Wagg}, J. 2019,
  arXiv e-prints, arXiv:1912.12699

\bibitem[{{Castor} {et~al.}(1975){Castor}, {Abbott}, \& {Klein}}]{Castor1975}
{Castor}, J.~I., {Abbott}, D.~C., \& {Klein}, R.~I. 1975, \apj, 195, 157

\bibitem[{{Das} {et~al.}(2022){Das}, {Chandra}, {Shultz}, {Leto},
  {Mikul{\'a}{\v{s}}ek}, {Petit}, \& {Wade}}]{Das_etal2022}
{Das}, B., {Chandra}, P., {Shultz}, M.~E., {et~al.} 2022, \mnras, 517, 5756

\bibitem[{{Dionne} \& {Robert}(2006)}]{Dionne2006}
{Dionne}, D. \& {Robert}, C. 2006, \apj, 641, 252

\bibitem[{{Doughty} \& {Finlator}(2021)}]{Doughty2021}
{Doughty}, C. \& {Finlator}, K. 2021, \mnras, 505, 2207

\bibitem[{{Drake} {et~al.}(1987){Drake}, {Abbott}, {Bastian}, {Bieging},
  {Churchwell}, {Dulk}, \& {Linsky}}]{Drake1987}
{Drake}, S.~A., {Abbott}, D.~C., {Bastian}, T.~S., {et~al.} 1987, \apj, 322,
  902

\bibitem[{{Drake} {et~al.}(1994){Drake}, {Linsky}, {Schmitt}, \&
  {Rosso}}]{Drake1994}
{Drake}, S.~A., {Linsky}, J.~L., {Schmitt}, J.~H.~M.~M., \& {Rosso}, C. 1994,
  \apj, 420, 387

\bibitem[{{Drout} {et~al.}(2023){Drout}, {G{\"o}tberg}, {Ludwig}, {Groh}, {de
  Mink}, {O'Grady}, \& {Smith}}]{Drout2023}
{Drout}, M.~R., {G{\"o}tberg}, Y., {Ludwig}, B.~A., {et~al.} 2023, arXiv
  e-prints, arXiv:2307.00061

\bibitem[{{Evans} {et~al.}(2011){Evans}, {DeGioia-Eastwood}, {Gagn{\'e}},
  {Townsley}, {Broos}, {Wolk}, {Naz{\'e}}, {Corcoran}, {Oskinova}, {Moffat},
  {Wang}, \& {Walborn}}]{Evans2011_lateB_no_Xray_sources}
{Evans}, N.~R., {DeGioia-Eastwood}, K., {Gagn{\'e}}, M., {et~al.} 2011, \apjs,
  194, 13

\bibitem[{{Gilkis} \& {Shenar}(2023)}]{Gilkis2023}
{Gilkis}, A. \& {Shenar}, T. 2023, \mnras, 518, 3541

\bibitem[{{G{\"o}tberg} {et~al.}(2018){G{\"o}tberg}, {de Mink}, {Groh},
  {Kupfer}, {Crowther}, {Zapartas}, \& {Renzo}}]{Goetberg2018}
{G{\"o}tberg}, Y., {de Mink}, S.~E., {Groh}, J.~H., {et~al.} 2018, \aap, 615,
  A78

\bibitem[{{Groh} {et~al.}(2008){Groh}, {Oliveira}, \& {Steiner}}]{Groh2008}
{Groh}, J.~H., {Oliveira}, A.~S., \& {Steiner}, J.~E. 2008, \aap, 485, 245

\bibitem[{{Gudennavar} {et~al.}(2012){Gudennavar}, {Bubbly}, {Preethi}, \&
  {Murthy}}]{Gudennavar2012}
{Gudennavar}, S.~B., {Bubbly}, S.~G., {Preethi}, K., \& {Murthy}, J. 2012,
  \apjs, 199, 8

\bibitem[{{Hamann} {et~al.}(2019){Hamann}, {Gr{\"a}fener}, {Liermann},
  {Hainich}, {Sander}, {Shenar}, {Ramachandran}, {Todt}, \&
  {Oskinova}}]{Hamann2019}
{Hamann}, W.~R., {Gr{\"a}fener}, G., {Liermann}, A., {et~al.} 2019, \aap, 625,
  A57

\bibitem[{{Hiltner} \& {Schild}(1966)}]{Hiltner1966}
{Hiltner}, W.~A. \& {Schild}, R.~E. 1966, \apj, 143, 770

\bibitem[{{Krti{\v{c}}ka} {et~al.}(2019){Krti{\v{c}}ka}, {Mikul{\'a}{\v{s}}ek},
  {Henry}, {Jan{\'\i}k}, {Kochukhov}, {Pigulski}, {Leto}, {Trigilio},
  {Krti{\v{c}}kov{\'a}}, {L{\"u}ftinger}, {Prv{\'a}k}, \&
  {Tich{\'y}}}]{Krt2019}
{Krti{\v{c}}ka}, J., {Mikul{\'a}{\v{s}}ek}, Z., {Henry}, G.~W., {et~al.} 2019,
  \aap, 625, A34

\bibitem[{{Kurapati} {et~al.}(2017){Kurapati}, {Chandra}, {Wade}, {Cohen},
  {David-Uraz}, {Gagne}, {Grunhut}, {Oksala}, {Petit}, {Shultz}, {Sundqvist},
  {Townsend}, \& {ud-Doula}}]{Kurapati2017}
{Kurapati}, S., {Chandra}, P., {Wade}, G., {et~al.} 2017, \mnras, 465, 2160

\bibitem[{{Leone} {et~al.}(1994){Leone}, {Trigilio}, \&
  {Umana}}]{Leone_etal1994}
{Leone}, F., {Trigilio}, C., \& {Umana}, G. 1994, \aap, 283, 908

\bibitem[{{Leto} {et~al.}(2022){Leto}, {Oskinova}, {Buemi}, {Shultz},
  {Cavallaro}, {Trigilio}, {Umana}, {Fossati}, {Pillitteri}, {Krti{\v{c}}ka},
  {Ignace}, {Bordiu}, {Bufano}, {Catanzaro}, {Cerrigone}, {Giarrusso},
  {Ingallinera}, {Loru}, {Owocki}, {Postnov}, {Riggi}, {Robrade}, \&
  {Leone}}]{Leto2022}
{Leto}, P., {Oskinova}, L.~M., {Buemi}, C.~S., {et~al.} 2022, \mnras, 515, 5523

\bibitem[{{Leto} {et~al.}(2020{\natexlab{a}}){Leto}, {Trigilio}, {Buemi},
  {Leone}, {Pillitteri}, {Fossati}, {Cavallaro}, {Oskinova}, {Ignace},
  {Krti{\v{c}}ka}, {Umana}, {Catanzaro}, {Ingallinera}, {Bufano}, {Riggi},
  {Cerrigone}, {Loru}, {Schillir{\'o}}, {Agliozzo}, {Phillips}, {Giarrusso}, \&
  {Robrade}}]{Leto2020b}
{Leto}, P., {Trigilio}, C., {Buemi}, C.~S., {et~al.} 2020{\natexlab{a}},
  \mnras, 499, L72

\bibitem[{{Leto} {et~al.}(2012){Leto}, {Trigilio}, {Buemi}, {Leone}, \&
  {Umana}}]{Leto2012}
{Leto}, P., {Trigilio}, C., {Buemi}, C.~S., {Leone}, F., \& {Umana}, G. 2012,
  \mnras, 423, 1766

\bibitem[{{Leto} {et~al.}(2006){Leto}, {Trigilio}, {Buemi}, {Umana}, \&
  {Leone}}]{Leto2006}
{Leto}, P., {Trigilio}, C., {Buemi}, C.~S., {Umana}, G., \& {Leone}, F. 2006,
  \aap, 458, 831

\bibitem[{{Leto} {et~al.}(2021){Leto}, {Trigilio}, {Krti{\v{c}}ka}, {Fossati},
  {Ignace}, {Shultz}, {Buemi}, {Cerrigone}, {Umana}, {Ingallinera}, {Bordiu},
  {Pillitteri}, {Bufano}, {Oskinova}, {Agliozzo}, {Cavallaro}, {Riggi}, {Loru},
  {Todt}, {Giarrusso}, {Phillips}, {Robrade}, \& {Leone}}]{Leto2021}
{Leto}, P., {Trigilio}, C., {Krti{\v{c}}ka}, J., {et~al.} 2021, \mnras, 507,
  1979

\bibitem[{{Leto} {et~al.}(2020{\natexlab{b}}){Leto}, {Trigilio}, {Leone},
  {Pillitteri}, {Buemi}, {Fossati}, {Cavallaro}, {Oskinova}, {Ignace},
  {Krti{\v{c}}ka}, {Umana}, {Catanzaro}, {Ingallinera}, {Bufano}, {Agliozzo},
  {Phillips}, {Cerrigone}, {Riggi}, {Loru}, {Munari}, {Gangi}, {Giarrusso}, \&
  {Robrade}}]{Leto2020}
{Leto}, P., {Trigilio}, C., {Leone}, F., {et~al.} 2020{\natexlab{b}}, \mnras,
  493, 4657

\bibitem[{{Leto} {et~al.}(2017){Leto}, {Trigilio}, {Oskinova}, {Ignace},
  {Buemi}, {Umana}, {Ingallinera}, {Todt}, \& {Leone}}]{Leto2017}
{Leto}, P., {Trigilio}, C., {Oskinova}, L., {et~al.} 2017, \mnras, 467, 2820

\bibitem[{{Leto} {et~al.}(2018){Leto}, {Trigilio}, {Oskinova}, {Ignace},
  {Buemi}, {Umana}, {Ingallinera}, {Leone}, {Phillips}, {Agliozzo}, {Todt}, \&
  {Cerrigone}}]{Leto2018}
{Leto}, P., {Trigilio}, C., {Oskinova}, L.~M., {et~al.} 2018, \mnras, 476, 562

\bibitem[{{Linsky} {et~al.}(1992){Linsky}, {Drake}, \& {Bastian}}]{Linsky1992}
{Linsky}, J.~L., {Drake}, S.~A., \& {Bastian}, T.~S. 1992, \apj, 393, 341

\bibitem[{{Naz{\'e}} {et~al.}(2014){Naz{\'e}}, {Petit}, {Rinbrand}, {Cohen},
  {Owocki}, {ud-Doula}, \& {Wade}}]{Naze2014}
{Naz{\'e}}, Y., {Petit}, V., {Rinbrand}, M., {et~al.} 2014, \apjs, 215, 10

\bibitem[{{Nebot G{\'o}mez-Mor{\'a}n} \& {Oskinova}(2018)}]{Nebot2018}
{Nebot G{\'o}mez-Mor{\'a}n}, A. \& {Oskinova}, L.~M. 2018, \aap, 620, A89

\bibitem[{{Neubauer} \& {Aller}(1948)}]{Neubauer1948}
{Neubauer}, F.~J. \& {Aller}, L.~H. 1948, \apj, 107, 281

\bibitem[{{Oskinova} {et~al.}(2012){Oskinova}, {Gayley}, {Hamann},
  {Huenemoerder}, {Ignace}, \& {Pollock}}]{Osk2012}
{Oskinova}, L.~M., {Gayley}, K.~G., {Hamann}, W.~R., {et~al.} 2012, \apjl, 747,
  L25

\bibitem[{{Oskinova} {et~al.}(2020){Oskinova}, {Gvaramadze}, {Gr{\"a}fener},
  {Langer}, \& {Todt}}]{Osk2020}
{Oskinova}, L.~M., {Gvaramadze}, V.~V., {Gr{\"a}fener}, G., {Langer}, N., \&
  {Todt}, H. 2020, \aap, 644, L8

\bibitem[{{Oskinova} {et~al.}(2011){Oskinova}, {Todt}, {Ignace}, {Brown},
  {Cassinelli}, \& {Hamann}}]{Osk2011}
{Oskinova}, L.~M., {Todt}, H., {Ignace}, R., {et~al.} 2011, \mnras, 416, 1456

\bibitem[{{Owocki} {et~al.}(2022){Owocki}, {Shultz}, {ud-Doula}, {Chandra},
  {Das}, \& {Leto}}]{Owocki2022}
{Owocki}, S.~P., {Shultz}, M.~E., {ud-Doula}, A., {et~al.} 2022, \mnras, 513,
  1449

\bibitem[{{Owocki} {et~al.}(2016){Owocki}, {ud-Doula}, {Sundqvist}, {Petit},
  {Cohen}, \& {Townsend}}]{Owocki2016}
{Owocki}, S.~P., {ud-Doula}, A., {Sundqvist}, J.~O., {et~al.} 2016, \mnras,
  462, 3830

\bibitem[{{Panagia} \& {Felli}(1975)}]{Panagia1975}
{Panagia}, N. \& {Felli}, M. 1975, \aap, 39, 1

\bibitem[{{Petit} {et~al.}(2015){Petit}, {Cohen}, {Wade}, {Naz{\'e}}, {Owocki},
  {Sundqvist}, {ud-Doula}, {Fullerton}, {Leutenegger}, \&
  {Gagn{\'e}}}]{Petit2015}
{Petit}, V., {Cohen}, D.~H., {Wade}, G.~A., {et~al.} 2015, \mnras, 453, 3288

\bibitem[{{Petit} {et~al.}(2013){Petit}, {Owocki}, {Wade}, {Cohen},
  {Sundqvist}, {Gagn{\'e}}, {Ma{\'\i}z Apell{\'a}niz}, {Oksala}, {Bohlender},
  {Rivinius}, {Henrichs}, {Alecian}, {Townsend}, {ud-Doula}, \& {MiMeS
  Collaboration}}]{Petit2013}
{Petit}, V., {Owocki}, S.~P., {Wade}, G.~A., {et~al.} 2013, \mnras, 429, 398

\bibitem[{{Podsiadlowski} {et~al.}(1992){Podsiadlowski}, {Joss}, \&
  {Hsu}}]{Podsi1992}
{Podsiadlowski}, P., {Joss}, P.~C., \& {Hsu}, J.~J.~L. 1992, \apj, 391, 246

\bibitem[{{Ramachandran} {et~al.}(2023){Ramachandran}, {Klencki}, {Sander},
  {Pauli}, {Shenar}, {Oskinova}, \& {Hamann}}]{Ramachandran2023}
{Ramachandran}, V., {Klencki}, J., {Sander}, A.~A.~C., {et~al.} 2023, \aap,
  674, L12

\bibitem[{{Ritter} {et~al.}(2018){Ritter}, {Herwig}, {Jones}, {Pignatari},
  {Fryer}, \& {Hirschi}}]{Ritter2018}
{Ritter}, C., {Herwig}, F., {Jones}, S., {et~al.} 2018, \mnras, 480, 538

\bibitem[{{Robrade} {et~al.}(2018){Robrade}, {Oskinova}, {Schmitt}, {Leto}, \&
  {Trigilio}}]{Robrade2018}
{Robrade}, J., {Oskinova}, L.~M., {Schmitt}, J.~H.~M.~M., {Leto}, P., \&
  {Trigilio}, C. 2018, \aap, 619, A33

\bibitem[{{Sander} {et~al.}(2019){Sander}, {Hamann}, {Todt}, {Hainich},
  {Shenar}, {Ramachandran}, \& {Oskinova}}]{Sander2019}
{Sander}, A.~A.~C., {Hamann}, W.~R., {Todt}, H., {et~al.} 2019, \aap, 621, A92

\bibitem[{{Scuderi} {et~al.}(1998){Scuderi}, {Panagia}, {Stanghellini},
  {Trigilio}, \& {Umana}}]{Scuderi1998}
{Scuderi}, S., {Panagia}, N., {Stanghellini}, C., {Trigilio}, C., \& {Umana},
  G. 1998, \aap, 332, 251

\bibitem[{{Selina} {et~al.}(2018){Selina}, {Murphy}, {McKinnon}, {Beasley},
  {Butler}, {Carilli}, {Clark}, {Durand}, {Erickson}, {Grammer}, {Hiriart},
  {Jackson}, {Kent}, {Mason}, {Morgan}, {Ojeda}, {Rosero}, {Shillue},
  {Sturgis}, \& {Urbain}}]{Selina2018}
{Selina}, R.~J., {Murphy}, E.~J., {McKinnon}, M., {et~al.} 2018, in
  Astronomical Society of the Pacific Conference Series, Vol. 517, Science with
  a Next Generation Very Large Array, ed. E.~{Murphy}, 15

\bibitem[{{Shenar} {et~al.}(2020){Shenar}, {Gilkis}, {Vink}, {Sana}, \&
  {Sander}}]{Shenar2020}
{Shenar}, T., {Gilkis}, A., {Vink}, J.~S., {Sana}, H., \& {Sander}, A.~A.~C.
  2020, \aap, 634, A79

\bibitem[{{Shenar} {et~al.}(2023){Shenar}, {Wade}, {Marchant}, {Bagnulo},
  {Bodensteiner}, {Bowman}, {Gilkis}, {Langer}, {Nicolas-Chen{\'e}},
  {Oskinova}, {Van Reeth}, {Sana}, {St-Louis}, {de Oliveira}, {Todt}, \&
  {Toonen}}]{Shenar2023}
{Shenar}, T., {Wade}, G.~A., {Marchant}, P., {et~al.} 2023, Science, 381, 761

\bibitem[{{Shultz} {et~al.}(2022){Shultz}, {Owocki}, {ud-Doula}, {Biswas},
  {Bohlender}, {Chandra}, {Das}, {David-Uraz}, {Khalack}, {Kochukhov},
  {Landstreet}, {Leto}, {Monin}, {Neiner}, {Rivinius}, \& {Wade}}]{Shultz2022}
{Shultz}, M.~E., {Owocki}, S.~P., {ud-Doula}, A., {et~al.} 2022, \mnras, 513,
  1429

\bibitem[{{Smith} {et~al.}(2001){Smith}, {Brickhouse}, {Liedahl}, \&
  {Raymond}}]{apec}
{Smith}, R.~K., {Brickhouse}, N.~S., {Liedahl}, D.~A., \& {Raymond}, J.~C.
  2001, \apjl, 556, L91

\bibitem[{{Stelzer} {et~al.}(2005){Stelzer}, {Flaccomio}, {Montmerle},
  {Micela}, {Sciortino}, {Favata}, {Preibisch}, \& {Feigelson}}]{Stelzer2005}
{Stelzer}, B., {Flaccomio}, E., {Montmerle}, T., {et~al.} 2005, \apjs, 160, 557

\bibitem[{{Trigilio} {et~al.}(2000){Trigilio}, {Leto}, {Leone}, {Umana}, \&
  {Buemi}}]{Trigilio2000}
{Trigilio}, C., {Leto}, P., {Leone}, F., {Umana}, G., \& {Buemi}, C. 2000,
  \aap, 362, 281

\bibitem[{{Trigilio} {et~al.}(2004){Trigilio}, {Leto}, {Umana}, {Leone}, \&
  {Buemi}}]{Trigilio_etal2004}
{Trigilio}, C., {Leto}, P., {Umana}, G., {Leone}, F., \& {Buemi}, C.~S. 2004,
  \aap, 418, 593

\bibitem[{{ud-Doula} \& {Naz{\'e}}(2016)}]{udD2016}
{ud-Doula}, A. \& {Naz{\'e}}, Y. 2016, Advances in Space Research, 58, 680

\bibitem[{{ud-Doula} {et~al.}(2014){ud-Doula}, {Owocki}, {Townsend}, {Petit},
  \& {Cohen}}]{ud-doula2014}
{ud-Doula}, A., {Owocki}, S., {Townsend}, R., {Petit}, V., \& {Cohen}, D. 2014,
  \mnras, 441, 3600

\bibitem[{{ud-Doula} \& {Owocki}(2002)}]{ud-doula2002}
{ud-Doula}, A. \& {Owocki}, S.~P. 2002, \apj, 576, 413

\bibitem[{{ud-Doula} {et~al.}(2008){ud-Doula}, {Owocki}, \&
  {Townsend}}]{ud-doula2008}
{ud-Doula}, A., {Owocki}, S.~P., \& {Townsend}, R. H.~D. 2008, \mnras, 385, 97

\bibitem[{{ud-Doula} {et~al.}(2013){ud-Doula}, {Sundqvist}, {Owocki}, {Petit},
  \& {Townsend}}]{ud-doula2013}
{ud-Doula}, A., {Sundqvist}, J.~O., {Owocki}, S.~P., {Petit}, V., \&
  {Townsend}, R.~H.~D. 2013, \mnras, 428, 2723

\bibitem[{{Usov} \& {Melrose}(1992)}]{Usov_Melrose1992}
{Usov}, V.~V. \& {Melrose}, D.~B. 1992, \apj, 395, 575

\bibitem[{{van Blerkom}(1978)}]{vanBlerkom1978}
{van Blerkom}, D. 1978, \apj, 225, 175

\bibitem[{{Vink}(2017)}]{Vink2017}
{Vink}, J.~S. 2017, \aap, 607, L8

\bibitem[{{Willis} {et~al.}(1989){Willis}, {Howarth}, {Stickland}, \&
  {Heap}}]{Willis1989}
{Willis}, A.~J., {Howarth}, I.~D., {Stickland}, D.~J., \& {Heap}, S.~R. 1989,
  \apj, 347, 413

\bibitem[{{Willis} \& {Stickland}(1983)}]{Willis1983}
{Willis}, A.~J. \& {Stickland}, D.~J. 1983, \mnras, 203, 619

\bibitem[{{Wilms} {et~al.}(2000){Wilms}, {Allen}, \& {McCray}}]{tbabs}
{Wilms}, J., {Allen}, A., \& {McCray}, R. 2000, \apj, 542, 914

\bibitem[{{Woosley} {et~al.}(1995){Woosley}, {Langer}, \&
  {Weaver}}]{Woosley1995}
{Woosley}, S.~E., {Langer}, N., \& {Weaver}, T.~A. 1995, \apj, 448, 315

\bibitem[{{Wright} \& {Barlow}(1975)}]{Wright1975}
{Wright}, A.~E. \& {Barlow}, M.~J. 1975, \mnras, 170, 41

\bibitem[{{Yungelson} {et~al.}(2024){Yungelson}, {Kuranov}, {Postnov},
  {Kuranova}, {Oskinova}, \& {Hamann}}]{Yung2024}
{Yungelson}, L., {Kuranov}, A., {Postnov}, K., {et~al.} 2024, \aap, 683, A37

\end{thebibliography}


\end{document}